\documentclass[a4paper]{ceurart}
\usepackage{listings}
\usepackage{graphicx}
\usepackage{adjustbox}
\usepackage{array}
\usepackage[caption=false]{subfig}
\usepackage{multicol}
\usepackage{float} 
\usepackage{hyperref}
\usepackage{enumitem}
\usepackage{soul}\setuldepth{article}
\usepackage{xcolor}
\usepackage{amsmath,graphicx}
\usepackage{pifont}
\newcommand{\cmark}{\ding{51}}
\newcommand{\xmark}{\ding{55}}
\usepackage{booktabs}
\usepackage{adjustbox}
\usepackage{tabularx}
\usepackage{makecell}
\usepackage{appendix}

\begin{document}

\copyrightyear{2024}
\copyrightclause{Copyright © 2024 for this paper by its authors. Use permitted under Creative Commons License Attribution 4.0 International (CC BY 4.0).}

\conference{International Workshop on Semantic Materials Science: Harnessing the Power of Semantic Web (SeMats), September 17--19, 2024, Amsterdam, Netherlands}

\title{The landscape of ontologies in materials science and engineering: A survey and evaluation}


\author[1]{Ebrahim Norouzi}[%
orcid=0000-0002-2691-6995,
email=ebrahim.norouzi@fiz-karlsruhe.de,
]
\cormark[1]
\address[1]{FIZ Karlsruhe – Leibniz-Institute for Information Infrastructure,
 Hermann-von-Helmholtz-Platz 1, 76344 Eggenstein-Leopoldshafen, Germany}

\author[1]{Jörg Waitelonis}[%
orcid=0000-0001-7192-7143,
]


\author[1,2]{Harald Sack}[%
orcid=0000-0001-7069-9804,
]

\address[2]{Karlsruhe Institute of Technology, Institute of Applied Informatics and Formal Description Methods, Kaiserstrasse 89, 76133 Karlsruhe, Germany}
\cortext[1]{Corresponding author.}

\begin{abstract}
Ontologies are widely used in materials science to describe experiments, processes, material properties, and experimental and computational workflows. Numerous online platforms are available for accessing and sharing ontologies in Materials Science and Engineering (MSE). Additionally, several surveys of these ontologies have been conducted. However, these studies often lack comprehensive analysis and quality control metrics. This paper provides an overview of ontologies used in Materials Science and Engineering to assist domain experts in selecting the most suitable ontology for a given purpose. Sixty selected ontologies are analyzed and compared based on the requirements outlined in this paper. Statistical data on ontology reuse and key metrics are also presented. The evaluation results provide valuable insights into the strengths and weaknesses of the investigated MSE ontologies. This enables domain experts to select suitable ontologies and to incorporate relevant terms from existing resources. 
\end{abstract}


\maketitle

\section{Introduction}
\label{sec:intro}

The field of Materials Science and Engineering (MSE) has seen a rapid increase in the number of available ontologies, taxonomies, terminologies, and vocabularies to semantically describe various aspects of research, such as experimental procedures, material properties, computational workflows, and experimental outcomes~\cite{zhangoverview}. Online repositories and portals such as MatPortal\footnote{\url{https://matportal.org/}}, BioPortal\footnote{\url{https://bioportal.bioontology.org/}}, and IndustryPortal\footnote{\url{http://industryportal.enit.fr/}} provide access to these MSE ontologies. However, a significant drawback is the relatively small number of registered and published ontologies. Moreover, the metadata accompanying many ontologies is often inadequate, hindering users' ability to understand the ontology's scope and domain. This makes it difficult for MSE domain experts to assess the ontology's relevance to their specific needs.

While several surveys have examined ontologies in the MSE domain, they often lack comprehensive coverage and in-depth analysis. For example, the survey by Zhang et al .\cite{zhangoverview} evaluates nine ontologies, six of which are currently inaccessible, hindering content verification. Similarly, Bayerlein et al. \cite{bernd_overview_2022} provide a general overview but fall short in offering comprehensive evaluations and quality assessments. Recently, Baas et al.  \cite{Gerhard_domain_os} reviewed 43 domain-level ontologies (DLOs) for Materials Science, providing an overview of their features and proposing an alignment methodology to enhance interoperability. While their study offers valuable insights, it lacks a comprehensive evaluation of ontology quality based on quality-control metrics. Additionally, these surveys, along with others, often overlook detailed analysis of usability, structural complexity, and application relevance, hindering domain experts in their ontology selection process.

Consequently, MSE domain experts struggle to identify and select suitable ontologies for their applications. A comprehensive review encompassing all existing ontologies, detailing their reuse in the MSE domain, and providing a quality assessment is urgently needed to guide expert decision-making.

This paper aims to address the current lack of comprehensive knowledge about MSE ontologies by providing a thorough review and analysis. Our objectives include identifying available ontologies, understanding their specific purposes, developing criteria for ontology selection, and determining the necessary information for informed decision-making by domain experts. All evaluation results are available online\footnote{\url{https://ise-fizkarlsruhe.github.io/mseo.github.io/}}.

The paper is structured as follows: Section~\ref{sec:method} describes the methodology used to collect and evaluate the MSE ontologies. Section~\ref{sec:results_discussion} presents quantitative results of the review process and provides a thorough discussion of these findings in the context of our study objectives. Finally, Section~\ref{sec:Conclusion_outlook} summarizes the primary outcomes of this study and outlines potential future research directions.

\section{Methodology}
\label{sec:method}

This section outlines the comprehensive methodology employed to evaluate ontologies in the Materials Science and Engineering (MSE) domain. The methodology comprises three Key components: expert insights and surveys, quality requirements and corresponding criteria, and ontology evaluation metrics. Each component significantly contributes to a thorough assessment of ontologies, empowering MSE experts to select the most suitable ontology for their specific needs. The following subsections provide detailed explanations of each methodological aspect:

\subsection{Expert Insights and Survey}
This subsection details the process of gathering expert insights and conducting a survey to identify ontology requirements within the MSE domain. Expert insights provide a foundational understanding of practical needs and challenges faced by domain professionals. To ensure the relevance and effectiveness of the selected ontologies, an internal survey was conducted as part of the Platform Material Digital (PMD) project\footnote{\url{https://www.materialdigital.de/}}. The PMD project comprises 13 industry-led pilot projects with the shared requirement of (re)using ontologies in the MSE domain. A key finding from our qualitative analysis is that the responses of the 13 projects focused on different aspects of the ontologies. This highlights the varied priorities and perspectives of the participating experts. The publicly available interview results\footnote{\url{https://git.material-digital.de/ontologies/pmd-ontologies/-/tree/main/Partner\%20project\%20CQs}} highlight the specific requirements identified by MSE experts. We asked the responsible PIs of the thirteen PMD projects to gather essential information about MSE-related ontologies within their domain. The survey targeted MSE experts, including material engineers, scientists, process and application engineers, and simulation experts, aiming to standardize taxonomy and metadata for materials and their properties. Key survey questions addressed the desired ontology domains, intended use and requirements of the ontology, target users, and the specific competency questions (CQ) the ontology should answer. Table~\ref{tab:quality_requirements} presents quality requirements, their justifications, and corresponding criteria, as determined by survey results. The criteria mapped to these requirements include Completeness (ensuring sufficient information for specified tasks), Coverage (measuring the breadth of domain information), Availability (assessing the accessibility of the ontology and its documentation), and Adaptability  (evaluating the ontology's capacity to accommodate changes without compromising verified definitions). Relevancy measures the ontology's alignment to specified tasks, while Accuracy assesses the precision and correctness of its representations. Compliance ensures adherence to defined rules and standards, and Internal Consistency guarantees logical coherence. Credibility reflects the ontology's acceptance and trustworthiness, and Complexity measures its structural intricacy. Finally, Comprehensibility measures users' understanding, and Modularity assesses the ontology's composability from discrete, manageable units \cite{wilson2022conceptual}. 

This paper specifically focuses on Availability (presence of ontology files and documentation), Adaptability (number of CQs answered correctly after changes), Complexity (structural properties like depth and breadth), Internal Consistency (presence of logical/formal contradictions), Compliance (total number of breached rules), Credibility (number of other ontologies that link to it or positive user feedback), Comprehensibility (degree of annotations and naming conventions), and Modularity (number of ontology partitions and root nodes). While the metrics for these criteria are defined, future work will address Coverage, Completeness, and Accuracy for a more holistic evaluation \cite{wilson2022conceptual}.

\subsection{Ontology Evaluation Metrics}
This subsection outlines the detailed metrics employed to evaluate the ontologies within the MSE domain. Focusing on base, schema, and graph metrics, we measure ontology structure, complexity, and usability. These metrics,
 categorized accordingly, were used in conjunction with the ROBOT tool\footnote{\url{http://robot.obolibrary.org/}} \cite{jackson2019robot} and OntoMetrics\footnote{\url{https://ontometrics.informatik.uni-rostock.de/ontologymetrics/}} \cite{lantow2016ontometrics} to assess ontology quality. 

Table~\ref{tab:metrics} presents the key metrics used in our evaluation including descriptions and significance. These metrics align with W3C Semantic Web standards, specifically OWL and OWL DL. \textbf{Base metrics} include Axioms, Class Count, Object Properties Count, Datatype Properties Count, Annotation Assertions Count, and DL Expressivity. These metrics collectively indicate the overall size, complexity, and breadth of the ontology, as well as the richness of relationships, data values, and applied logical constructs. For \textbf{schema metrics}, Attribute Richness, Inheritance Richness, Relationship Richness, Axiom Class Ratio, and Equivalence Ratio are included. These metrics reflect the detailed knowledge representation, categorization, interconnectedness, logical definition detail, and redundancy among named classes. Complex class definitions are excluded from these metrics. \textbf{Graph metrics} evaluated include Absolute Root Cardinality (NoR), Absolute Leaf Cardinality (NoL), Number of External Classes (NoC), Depth, Breadth, and Tangledness are evaluated. These metrics assess the foundational structure, granularity, interdependence with external ontologies, hierarchical complexity, width, and interconnectivity of the ontology.

Furthermore, the OOPS! (Ontology Pitfall Scanner!) tool \cite{poveda2014oops} was to detect common ontology development errors, including missing property domains or ranges, incorrect subclassing, and redundant relationships. Identifying these pitfalls is crucial for ensuring ontology usability and effectiveness in real-world applications. A detailed overview of the identified issues of MSE ontologies is provided in the Appendix~\ref{appendix_pitfalls}.

\begin{table}[t]
\centering
\caption{Quality requirements and their justifications with mapped criteria.}
\label{tab:quality_requirements}
\begin{adjustbox}{width=\textwidth}
\begin{tabular}{m{6.5cm}|m{5.5cm}|m{3cm}}
\hline
\textbf{Quality Requirements} & \textbf{Justification} & \textbf{Mapped Criteria} \\
\hline
REQ1: The ontology must provide a comprehensive taxonomy covering necessary sub-domains of materials science. & The term "comprehensive" suggests that the ontology should be complete and cover all necessary sub-domains. & Completeness, Coverage, Availability \\
\hline
REQ2: The ontology should be capable of representing both experimental and simulation data. & The capability to represent various types of data indicates a complete and adaptable ontology. & Completeness, Adaptability, Availability \\
\hline
REQ3: The ontology should accurately represent relations between different concepts. & Accuracy is crucial for representing relationships, and relevancy ensures that these relationships matter to the domain. & Accuracy, Relevancy, Availability \\
\hline
REQ4: The ontology must standardize and unify metadata descriptions in compliance with existing standards. & Standardization implies compliance and internal consistency. & Compliance, Internal Consistency, Availability \\
\hline
REQ5: The ontology must provide comprehensive representations for experimental settings, outcomes, high-throughput analysis data, and literature. & Comprehensive representations imply completeness and coverage. & Completeness, Coverage, Availability \\
\hline
REQ6: The ontology must facilitate trustworthy and verifiable quality management of data. & Trustworthiness and verifiability imply credibility and accuracy. & Credibility, Accuracy, Availability \\
\hline
REQ7 (more specific): The ontology should enable querying specifically designed for machine learning model development. & Specificity for machine learning makes it relevant, complex, and comprehensive for that purpose. & Relevancy, Complexity, Comprehensibility, Availability \\
\hline
REQ8 (more specific): The ontology must represent predicted values from machine learning models. & Representation of machine learning and predicted properties implies completeness and coverage. & Completeness, Coverage, Availability \\
\hline
REQ9: The ontology should be modular enough for diverse projects beyond its primary application. & Modularity and the ability to be used in diverse projects imply adaptability. & Adaptability, Modularity, Availability \\
\hline
\end{tabular}
\end{adjustbox}
\end{table}

\begin{table}
\centering
\caption{Summary of Ontology Evaluation Metrics (Structural Metrics) \cite{lantow2016ontometrics}.}
\label{tab:metrics}
\begin{adjustbox}{width=.9\textwidth}
\begin{tabular}{>{\centering\arraybackslash}m{.6cm}|m{2cm}|m{4cm}|m{8cm}}
\hline
\textbf{Type} & \textbf{Metric} & \textbf{Description} & \textbf{Impact} \\
\hline
\multirow{7}{*}{\rotatebox[origin=c]{90}{\parbox{6cm}{\centering Base Metrics}}}
& Axioms & Total number of axioms for classes, properties, datatype definitions, assertions, and annotations. & Indicates the overall size and complexity of the ontology. Higher numbers suggest richer, more comprehensive ontologies. \\
\cline{2-4}
& Class Count & Total number of classes defined in the ontology. & Shows the breadth of concepts covered. Higher counts imply broader domain coverage and the ability to represent a wide range of concepts. \\
\cline{2-4}
& Object Properties Count & Total number of object properties in the ontology. & Indicates the variety of relationships between classes. More object properties suggest richer interconnections and better representation of complex interactions. \\
\cline{2-4}
& Datatype Properties Count & Total number of datatype properties in the ontology. & Reflects the range of data values associated with classes. Higher counts indicate detailed representation of class attributes. \\
\cline{2-4}
& Annotation Assertions Count & Total number of annotations in the ontology. & Shows the amount of descriptive metadata provided. More annotations enhance documentation and understandability. \\
\cline{2-4}
& DL Expressivity & The description logic expressivity of the ontology. & Indicates the complexity of logical constructs used. Higher expressivity allows for nuanced and complex reasoning. \\
\hline
\multirow{5}{*}{\rotatebox[origin=c]{90}{\parbox{6cm}{\centering Schema Metrics}}}
& Attribute Richness & Average number of attributes per class. & Higher values indicate detailed knowledge representation, essential for conveying complex information and improving usability in applications. \\
\cline{2-4}
& Inheritance Richness & Average number of subclasses per class. & Helps in understanding how knowledge is categorized and structured. Higher values suggest better categorization and hierarchical structuring. \\
\cline{2-4}
& Relationship Richness & Ratio of non-inheritance relationships to total relationships. & Important for understanding interconnectedness. Higher values imply a more interconnected and informative ontology. \\
\cline{2-4}
& Axiom Class Ratio & Average number of axioms per class. & Indicates the level of detail in logical definitions. Higher ratios suggest more detailed and well-defined concepts. \\
\cline{2-4}
& Equivalence Ratio & Proportion of equivalence axioms to all classes. & Reflects redundancy and synonymy, indicating how terms are defined as equivalent. Higher values improve the ontology's integration ability. \\
\hline
\multirow{6}{*}{\rotatebox[origin=c]{90}{\parbox{8cm}{\centering Graph Metrics}}}
& Absolute Root Cardinality (NoR) & Number of root classes in the ontology. & Indicates cohesion and foundational structure. More root classes suggest a diversified foundational structure. \\
\cline{2-4}
& Absolute Leaf Cardinality (NoL) & Number of leaf classes in the ontology. & Assesses granularity and detail. More leaf classes enhance the ability to capture fine-grained knowledge. \\
\cline{2-4}
& Number of External Classes (NoC) & The total number of classes referenced from external ontologies. & Indicates the degree of interdependence with other ontologies. A higher count suggests better interoperability, reusability, and alignment with external standards, though it may also introduce complexity. \\ \cline{2-4}
& Depth & Depth from root to leaf nodes. & Shows hierarchical complexity. Greater depth indicates more detailed hierarchical levels. \\
\cline{2-4}
& Breadth & Breadth at each level, summed across all levels. & Indicates the number of sibling classes, reflecting the ontology's width. Greater breadth suggests comprehensive coverage at each level. \\
\cline{2-4}
& Tangledness & Degree of interconnectivity, measuring multiple hierarchical paths. & Evaluates complexity and overlap in the hierarchical structure. Higher tangledness suggests greater complexity and potential difficulty in navigation. \\
\hline
\end{tabular}
\end{adjustbox}
\end{table}

\begin{table}[t!]
\centering
\caption{List of Top, Middle, and Domain level ontologies in the domain of Materials Science and Engineering. The columns are defined as follows: Type (ontology level), Ontology Name, Domain, Used in Project(s), Competency Questions (CQs), Licensing (Lic.), Top-Level Ontology Alignment (TLO), Reused Ontologies (Reused Ont.), Modularity (Mod.), and Adoption of Ontology Design Patterns (ODPs). The list was updated in June, and all the ontologies were found to the best of our knowledge. \xmark  means that the information could not be found either from the ontology repository or from the publication reference of the ontology.}
\label{tab:general-1}
\begin{adjustbox}{width=\textwidth}
\begin{tabular}{m{0.7cm}|m{5cm}>{\centering\arraybackslash}m{3cm}>{\centering\arraybackslash}m{5cm}>{\centering\arraybackslash}m{1cm}>{\centering\arraybackslash}m{1cm}>{\centering\arraybackslash}m{1cm}>{\centering\arraybackslash}m{1cm}>{\centering\arraybackslash}m{1cm}>{\centering\arraybackslash}m{1cm}}
\hline
\textbf{Type} & \textbf{Ontology Name} & \textbf{Domain} & \textbf{Used in Project(s)} & \textbf{CQs} & \textbf{Lic.} & \textbf{TLO} & \textbf{Reused Ont.} & \textbf{Mod.} & \textbf{ODPs}   \\
\hline
\multirow{4}{*}{\rotatebox[origin=c]{90}{\parbox{7cm}{\centering Top-level}}} & Basic Formal Ontology (BFO) \cite{bfo} & Supporting information retrieval, analysis and integration in scientific and other domains & 300 ontologies, 50 organizations, PubChem, ODE\_AM, DIGITRUBBER & \xmark & CC BY 4.0 & \xmark & \xmark & \cmark & \xmark \\ \cline{2-10}
& Elementary Multiperspective Material Ontology (EMMO) \cite{emmo} & A multidisciplinary ontology for applied sciences & EMMC-CSA, SimDOME, MarketPlace, VIMMP, OntoTrans, ReaxPro, OntoCommons, OYSTER, NanoMECommons, OpenModel, Know-Now, iBain, KupferDigital, StahlDigital, SmaDi, DiProgMag, SensoTwin & \xmark & CC BY 4.0 & \xmark & \xmark & \cmark & \xmark \\ \cline{2-10}
& Suggested Upper Merged Ontology (SUMO) \cite{sumo} & An upper ontology designed to serve as a foundational framework for various computer information processing systems. & Adimen-SUMO, YAGO-SUMO integration, smart city initiatives, psychoinformatics, and many other projects. & \xmark & GNU Public License & \xmark & \xmark & \cmark & \xmark \\ \cline{2-10}
& Descriptive Ontology for Linguistic and Cognitive Engineering (DOLCE) \cite{dolce} & A multidisciplinary ontology for applied sciences & EBRiO, MITE, SMARTEST, SORTT, I-TROPHYTS, and many other projects. & \xmark & GNU Public License & \xmark & \xmark & \cmark & \xmark \\ \hline

\multirow{8}{*}{\rotatebox[origin=c]{90}{\parbox{8cm}{\centering Mid-level}}} & PMD Core Ontology (PMDCO) \cite{pmdco} & Materials Science and Experiment & Know-Now, KupferDigital, StahlDigital, DiProgMag, DigiBatMat, GlasDigital, SensoTwin, LeBeDigital & \cmark & \xmark & PROVO \cite{prov-o} & PROVO & \cmark & \xmark \\ \cline{2-10}
& Material Science and Engineering Ontology (MSEO) \cite{mseo} & Material Science and Engineering & Materials Open Laboratory, LeBeDigital, KupferDigital & \xmark & \xmark & BFO & \xmark & \xmark & \xmark \\ \cline{2-10}
& Baden Württemberg Material Digital Domain Mid Level Ontology (BWMD-MID) \cite{bwmd_review} & Materials Science & Materials Open Laboratory, AluTrace, DMD4Future, KMU-akut: ADAM & \xmark & \xmark & BFO & \xmark & \xmark & \xmark \\ \cline{2-10}
& EMMO Datamodel ontology (EMMO Datamodel) \cite{emmo_datamodel} & Ontology-Based Data Modelling & MarketPlace, OntoTrans, OpenModel, VIPCOAT & \xmark & CC BY 4.0 & EMMO & EMMO & \cmark & \xmark \\ \cline{2-10}
& Materials Data Science Ontology (MDS) \cite{mds-o} & Materials Science & Materials Data Science for Stockpile Stewardship: Center of Excellence, SDLE Research Center & \xmark & \xmark & BFO & PMDCO & \xmark & \xmark \\ \cline{2-10}
& Ontology of Scientific Experiments (EXPO) \cite{expo} & Scientific Experiments & - & \xmark & \xmark & SUMO & \xmark & \xmark & \xmark \\ \cline{2-10}
& The Open Provenance Model for Workflows (OPMW) \cite{OPMW} & Process Modeling & - & \xmark & CC BY-NC-SA 2.0 & PROVO & P-Plan \cite{p-plan} & \xmark & \xmark \\ \cline{2-10}
& EMMO Mappings ontology (EMMO Mappings) \cite{emmo-mapping-o} & Mapping to domains and ontological concepts & MarketPlace, OntoTrans, OpenModel & \xmark & CC BY 4.0 & EMMO & EMMO & \xmark & \xmark \\ \hline

\multirow{9}{*}{\rotatebox[origin=c]{90}{\parbox{7cm}{\centering Domain-level}}} & PRovenance Information in MAterials science (PRIMA) \cite{PRIMA} & An ontology that captures the provenance information in the materials science domain. & EOSC-Pillar project, Helmholtz Metadata Collaboration (HMC), NFFA-Europe Pilot & \cmark & CC BY 3.0 & PROVO & PROVO, QUDT \cite{qudt}, PMDCO & \cmark & \xmark \\ \cline{2-10}
& EMMO Mechanical Testing (EMMO Mechanical Testing) \cite{emmo-mechanical-o} & Mechanical testing & MarketPlace, Oyster, UrWerk, iBain, StahlDigital & \xmark & CC BY 4.0 & EMMO & EMMO & \xmark & \xmark \\ \cline{2-10}
& EMMO Microstructure (EMMO Microstructure) \cite{emmo-microstructure-o} & Metallic microstructures & iBain, StahlDigital & \xmark & \xmark & EMMO & EMMO & \xmark & \xmark \\ \cline{2-10}
& SMART-Protocols (SP) \cite{sp_ontology} & Experimental protocols & \xmark & \cmark & CC BY 4.0 & BFO & NPO & \cmark & \xmark \\ \cline{2-10}
& Baden Württemberg Material Digital Domain Ontology (BWMD-DOMAIN) \cite{bwmd_review} & Material Digital Ontology, GlasDigital & Materials Open Laboratory, AluTrace, DMD4Future, KMU-akut: ADAM & \xmark & \xmark & BFO & BWMD-MID & \xmark & \xmark \\ \cline{2-10}
& MatOnto (MatOnto) \cite{MatOnto-o} & Materials discovery & \xmark & \xmark & MIT & DOLCE & DOLCE & \cmark & \xmark \\ \cline{2-10}
& NanoParticle Ontology (NPO) \cite{npo_ontology} & Cancer nanotechnology research & caB2B, eNanoMapper, nano-TAB & \xmark & BSD-3-Clause & BFO & \xmark & \xmark & \xmark \\ \cline{2-10}
& Smart Applications REFerence (SAREF) \cite{saref_ontology} & Energy & EEBus, SmartM2M & \xmark & BSD-3-Clause & \xmark & \xmark & \xmark & \xmark \\ \cline{2-10}
& Semantic Sensor Network Ontology (SSN) \cite{ssn_ontology} & Sensors & SENSEI, SmartProducts, SPITFIRE FP7, SemsorGrid4Env, Exalted, CSIRO & \xmark & OpenGIS & DOLCE & DOLCE & \cmark & \xmark \\
\hline
\end{tabular}
\end{adjustbox}
\end{table}

\begin{table}[t]
\centering
\caption{List of Domain level ontologies in Materials Science and Engineering. The columns are defined as follows: Type (ontology level), Ontology Name, Domain, Used in Project(s), Competency Questions (CQs), Licensing (Lic.), Top-Level Ontology Alignment (TLO), Reused Ontologies (Reused Ont.), Modularity (Mod.), and Ontology Design Patterns (ODPs). The list was updated in June, and all the ontologies were found to the best of our knowledge. \xmark  means that the information could not be found either from the ontology repository or from the publication reference of the ontology.}
\label{tab:general-2}
\begin{adjustbox}{width=\textwidth}
\begin{tabular}{m{0.7cm}|m{5cm}>{\centering\arraybackslash}m{3cm}>{\centering\arraybackslash}m{5cm}>{\centering\arraybackslash}m{1cm}>{\centering\arraybackslash}m{1cm}>{\centering\arraybackslash}m{1cm}>{\centering\arraybackslash}m{1cm}>{\centering\arraybackslash}m{1cm}>{\centering\arraybackslash}m{1cm}}
\hline
\textbf{Type} & \textbf{Ontology Name} & \textbf{Domain} & \textbf{Used in Project(s)} & \textbf{CQs} & \textbf{Lic.} & \textbf{TLO} & \textbf{Reused Ont.} & \textbf{Mod.} & \textbf{ODPs}   \\
\hline
\multirow{22}{*}{\rotatebox[origin=c]{90}{\parbox{25cm}{\centering Domain-level}}} & Characterisation Methodology Domain Ontology (CHAMEO) \cite{CHAMEO} & Materials Characterization & NanoMECommons, OYSTER, Big-Map, OntoTran & \xmark & CC BY 4.0 & EMMO & EMMO & \xmark & \xmark \\ \cline{2-10}
& NanoMine (NanoMine) \cite{nanomine_ontology} & Polymer nanocomposites & NanoMine & \xmark & CC BY 4.0 & SIO & \xmark & \xmark & \xmark \\ \cline{2-10}
& EMMO General Process Ontology (GPO) \cite{GPO} & Model processes & KIproBatt, OpenSemanticLab, Battery Value Chain Ontology & \xmark & CC BY 4.0 & EMMO & EMMO & \xmark & \xmark \\ \cline{2-10}
& EMMO Battery Value Chain Ontology (EMMO BVC) \cite{BVC} & Model processes along the Battery value chain & eLi /eLi-PLUS, KiProBatt & \xmark & CC BY 4.0 & EMMO & EMMO, GPO & \xmark & \xmark \\ \cline{2-10}
& EMMO Crystallography (EMMO Crystallography) \cite{EMMO-Crystallography} & Crystallography & Demystify ontologies, OntoTrans, MarketPlace, BIG-MAP & \xmark & CC BY 4.0 & EMMO & EMMO, CIF core & \xmark & \xmark \\ \cline{2-10}
& CIF Core Ontology (CIF-core) \cite{CIF-Core} & Crystallography & Demystify ontologies, OntoTrans, MarketPlace, BIG-MAP & \xmark & CC BY 4.0 & EMMO & EMMO & \xmark & \xmark \\ \cline{2-10}
& EMMO Atomistic and Electronic Modelling (EMMO Atomistic) \cite{EMMO-Atomistic} & Atomistic and electronic modelling & MarketPlace & \xmark & CC BY 4.0 & EMMO & EMMO & \xmark & \xmark \\ \cline{2-10}
& Materials Ontology (MATINFO) \cite{matinfo} & Exchanging Materials Information and Knowledge & Data exchange of AIST, NIMS, MatDB & \xmark & \xmark & \xmark & \xmark & \cmark & \xmark \\ \cline{2-10}
& Semantic Materials, Manufacturing, and Design (SEMMD) \cite{SEMMD} & Materials, Manufacturing & \xmark & \xmark & CC BY-SA 4.0 & BFO & QUDT & \xmark & \xmark \\ \cline{2-10}
& eNanoMapper (eNanoMapper) \cite{enanomapper} & Nanomaterial safety assessment & NanoSolveIT, NanoCommons, OpenRiskNet, eNanoMapper, NANoREG, NanoReg2, GRACIOUS & \xmark & CC-BY 3.0 & BFO & NPO & \xmark & \xmark \\ \cline{2-10}
& Dislocation Ontology (DISO) \cite{diso_ontology} & Defects in crystalline materials & European Research Council through the ERC Grant Agreement No. 759419 MuDiLingo, Helmholtz Metadata Collaboration (HMC) within the Hub Information at the Forschungszentrum Jülich & \cmark & MIT & EMMO & EMMO, QUDT & \xmark & \xmark \\ \cline{2-10}
& EMMO Battery Interface Ontology (EMMO BattINFO) \cite{BattINFO} & Batteries and their interfaces & BIG-MAP & \xmark & CC BY 4.0 & EMMO & \xmark & \xmark & \xmark \\ \cline{2-10}
& MatWerk Ontology (MWO) \cite{mwo} & Research Data Description & NFDI MatWerk & \cmark & CC BY 4.0 & BFO & NFDIcore & \cmark & \xmark \\ \cline{2-10}
& Materials And Molecules Basic Ontology (MAMBO) \cite{mambo_ontology} & Molecular materials & \xmark & \xmark & \xmark & \xmark & MDO, EMMO, QUDT & \cmark & \xmark \\ \cline{2-10}
& Laser Powder Bed Fusion Ontology (LPBFO) \cite{lpbf_ontology} & Additive manufacturing & Materials Open Laboratory, AluTrace, DMD4Future, KMU-akut: ADAM & \xmark & \xmark & BFO & BWMD-MID & \xmark & \xmark \\ \cline{2-10}
& Additive Manufacturing Ontology (AMONTOLOGY) \cite{AMOntology} & Additive Manufacturing & NIST's Systems Integration for Additive Manufacturing project & \xmark & \xmark & \xmark & \xmark & \xmark & \xmark \\ \cline{2-10}
& Building Material Ontology (BMO) \cite{Building-Material} & Materials Science & Linked data and Ontologies for Semantic Interoperability, BIM4EEB & \xmark & CC BY 4.0 & \xmark & \xmark & \xmark & \xmark \\ \cline{2-10}
& Standards-Specific Ontology Standard (SSOS) \cite{ssos_ontology} & Life cycle information & \xmark & \xmark & \xmark & \xmark & \xmark & \xmark & \xmark \\ \cline{2-10}
& Materials Design Ontology (MDO) \cite{mdo} & Materials design & Materials Design (OPTIMADE) project & \cmark & MIT & EMMO & EMMO, PROVO & \cmark & \xmark \\ \cline{2-10}
& The Devices, Experimental scaffolds and Biomaterials Ontology (DEB) \cite{deb_ontology} & Medical devices, experimental scaffolds and biomaterials & Database of Experimental Biomaterials and their Biological Effect & \xmark & GPL-3.0 & \xmark & \xmark & \xmark & \xmark \\ \cline{2-10}
& Ontology for Simulation, Modelling, and Optimization (OSMO) \cite{OSMO} & Materials modeling and simulation & VIMMP Marketplace & \xmark & LGPL v3 & EMMO & EMMO & \xmark & \xmark \\ \cline{2-10}
& Materials Vocabulary (MatVoc) \cite{MatVoc} & Materials Science & STREAM project & \xmark & \xmark & \xmark & \xmark & \xmark & \xmark \\ \hline
\end{tabular}
\end{adjustbox}
\end{table}

\begin{table}[t]
\centering
\caption{List of Domain and Application level ontologies in Materials Science and Engineering. The columns are defined as follows: Type (ontology level), Ontology Name, Domain, Used in Project(s), Competency Questions (CQs), Licensing (Lic.), Top-Level Ontology Alignment (TLO), Reused Ontologies (Reused Ont.), Modularity (Mod.), and Ontology Design Patterns (ODPs).}
\label{tab:general-3}
\begin{adjustbox}{width=\textwidth}
\begin{tabular}{m{0.7cm}|m{5cm}>{\centering\arraybackslash}m{3cm}>{\centering\arraybackslash}m{5cm}>{\centering\arraybackslash}m{1cm}>{\centering\arraybackslash}m{1cm}>{\centering\arraybackslash}m{1cm}>{\centering\arraybackslash}m{1.5cm}>{\centering\arraybackslash}m{1cm}>{\centering\arraybackslash}m{1cm}}
\hline
\textbf{Type} & \textbf{Ontology Name} & \textbf{Domain} & \textbf{Used in Project(s)} & \textbf{CQs} & \textbf{Lic.} & \textbf{TLO} & \textbf{Reused Ont.} & \textbf{Mod.} & \textbf{ODPs}   \\
\hline
\multirow{24}{*}{\rotatebox[origin=c]{90}{\parbox{19cm}{\centering Domain-level}}} &Computational Material Sample Ontology (CMSO) & Computational Materials Science & NFDI MatWerk & \cmark & \xmark & \xmark & \xmark & \xmark & \xmark \\ \cline{2-10}
& Open Energy Ontology (OEO) \cite{oeo_ontology} & Energy modelling domain & SzenarienDB & \cmark & CC0 1.0 & BFO & IAO \cite{IAO}, RO \cite{RO}, UO \cite{UO}, OMO \cite{OMO-ontology} & \cmark & \xmark \\ \cline{2-10}
&tribAIn Ontology (tribAIn) \cite{tribAIn} & tribology & \xmark & \xmark & CC BY 4.0 & SUMO & \xmark & \xmark & \xmark \\ \cline{2-10}
& Chemical Methods Ontology (CHMO) \cite{CMO} & Materials Characterization & \xmark & \xmark & CC BY 4.0 & \xmark & BFO & \xmark & \xmark \\ \cline{2-10}
& Chemical Information Ontology (CHEMINF) \cite{CIO} & Chemistry & \xmark & \xmark & CC BY 3.0 & \xmark & BFO & \xmark & \xmark \\ \cline{2-10}
& ONTORULE steel ontology (ONTORULE) \cite{ONTORULE} & Steel Industry & ONTORULE project & \xmark & \xmark & \xmark & \xmark & \xmark & \xmark \\ \cline{2-10}
& Photovoltaics Ontology (Photovoltaics) \cite{Photovoltaics} & Perovskite Solar Cells & \xmark & \xmark & CC BY 4.0 & \xmark & EMMO & \xmark & \xmark \\ \cline{2-10}
& Materials Data Science Ontology (MDS) \cite{MDS} & Materials Science & Materials Data Science for Stockpile Stewardship: Center of Excellence, SDLE Research Center & \xmark & \xmark & BFO & PMDCO, BFO & \xmark & \xmark \\ \cline{2-10}
& Dislocation Simulation and Model Ontology (DSIM) \cite{DSIM} & Computational Materials Science & NFDI-MatWerk & \xmark & CC BY 3.0 & \xmark & PROVO, MDO, QUDT, CSO, DISO & \xmark & \xmark \\ \cline{2-10}
& Crystal Structure Ontology (CSO) \cite{CSO} & Materials Characterization & MuDiLingo, Helmholtz Metadata Collaboration (HMC) & \cmark & CC BY 3.0 & \xmark & MDO, QUDT, ChEBI, BFO & \xmark & \xmark \\ \cline{2-10}
& Atomistic Simulation Methods Ontology (ASMO) \cite{ASMO} & Computational Materials Science & NFDI-MatWerk & \xmark & CC BY 4.0 & \xmark & PROVO, MDO & \xmark & \xmark \\ \cline{2-10}
& Point Defects Ontology (PODO) \cite{PODO} & Materials Characterization & NFDI-MatWerk & \xmark & CC BY 4.0 & \xmark & \xmark & \xmark & \xmark \\ \cline{2-10}
& Crystallographic Defect Core Ontology (CDCO) \cite{CDCO} & Materials Characterization & \xmark & \xmark & \xmark & \xmark & \xmark & \xmark & \xmark \\ \cline{2-10}
& Line Defect Ontology (LDO) \cite{LDO} & Materials Characterization & \xmark & \xmark & \xmark & \xmark & \xmark & \xmark & \xmark \\ \cline{2-10}
& Planar Defects Ontology (PLDO) \cite{PLDO} & Materials Characterization & NFDI-MatWerk & \xmark & CC BY 4.0 & \xmark & \xmark & \xmark & \xmark \\ \cline{2-10}
& Material Science Lab Equipment Ontology (MSLE) \cite{MSLE} & Materials Characterization & \xmark & \xmark & \xmark & \xmark & \xmark & \xmark & \xmark \\ \cline{2-10}
& Open Innovation Environment (OIE) - 5 Ontologies \cite{OIE-o} & Materials Science & OYSTER & \xmark & CC BY 4.0 & \xmark & EMMO & \cmark & \xmark \\ \cline{2-10}
& Metadata4Ing Ontology (M4I) \cite{M4I} & Process Modeling & NFDI4Ing & \cmark & CC BY 4.0 & \xmark & \xmark & \xmark & \xmark \\ \cline{2-10}
& MaterialsMine (MaterialsMine) \cite{MaterialsMine} & Materials Science & \xmark & \xmark & MIT & SIO & NanoMine, PROVO & \xmark & \xmark \\ \cline{2-10}
& Periodic Table Ontology (Periodictable) \cite{Periodictable} & Representation of the Periodic Table of the Elements & \xmark & \xmark & \xmark & \xmark & \xmark & \xmark & \xmark \\ \cline{2-10}
& Ontology to Describe Workflows in Linked Data (WILD) \cite{WILD} & Process Modeling & DigiBatMat & \xmark & \xmark & \xmark & FOAF & \xmark & \xmark \\ \cline{2-10}
& Ontology for Biomedical Investigations (OBI) \cite{OBI} & Representation of study design, protocols and instrumentation in Biomedicine & OBO Foundry, Allotrope™, PubChem & \xmark & CC-BY 3.0 & BFO & GO \cite{GO}, ChEBI, PATO, OBO & \xmark & \xmark \\ \cline{2-10}
& Phenotype And Trait Ontology (PATO) \cite{PATO} & Phenotypic, Physical Qualities & OBO Foundry, Allotrope™ & \xmark & CC-BY 4.0 & BFO & GO, ChEBI & \xmark & \xmark \\ \cline{2-10}
& Material properties ontology (MAT) \cite{Material-Properties} & Materials Science & H2020 BIMERR Project & \xmark & CC BY 4.0 & \xmark & SAREF & \xmark & \xmark \\ \hline 

\multirow{2}{*}{\rotatebox[origin=c]{90}{\parbox{1.5cm}{\centering Appl.\newline -level}}}

& Matolab Tensile Test Ontology (MOL TENSILE) \cite{MOL-TENSILE} & Tensile test & Materials Open Laboratory & \xmark & \xmark & \xmark & BWMD-MID & \xmark & \xmark \\ \cline{2-10}
& Matolab Brinell Test Ontology (MOL BRINELL) \cite{MOL-BRINELL} & Mechanical testing & \xmark & \xmark & \xmark & \xmark & \xmark & \xmark & \xmark \\
\hline
\end{tabular}
\end{adjustbox}
\end{table}

\section{Results and Discussion}
\label{sec:results_discussion}

\subsection{Analysis of the MSE Ontologies}
\label{sec:ontologies_results_section}

In our comprehensive analysis of semantic artifacts relevant to the Materials Science and Engineering (MSE) domain, a total of 94~semantic artifacts were identified, comprising 2~vocabularies and 92~ontologies. These ontologies can be categorized as follows: 11~general scientific ontologies, 7~without publicly available files, 4~top-level ontologies, 8~mid-level ontologies, 60~domain-level ontologies, and 2~application-level ontologies. Notably, 7~ontologies could not be evaluated due to issues with their imports. These ontologies include the Virtual Materials Marketplace (VIMMP) \cite{vimmp_ontology} Ontology, Chemical Entities of Biological Interest (ChEBI) \cite{chebi}, Chemical Information Ontology (CHEMINF) \cite{CHEMINF}, Metal Alloy (MetalAlloy) \cite{MetalAlloy}, tribAIn Ontology \cite{tribAIn}, Semantic Materials  Manufacturing Design (SEMMD) \cite{SEMMD}, and Chemical Methods Ontology (CHMO) \cite{CHAMEO}. This highlights the necessity for improved import handling and integration in ontology development processes. Consequently, 60~ontologies, as summarized in Tables~\ref{tab:general-1}, \ref{tab:general-2}, and \ref{tab:general-3}, were evaluated using our introduced methodology\footnote{\href{https://docs.google.com/spreadsheets/d/1I4Ye2lrAn68mrd9edq1m6CYx2DdWgDFx/edit?usp=sharing&ouid=100247787073779135889&rtpof=true&sd=true}{Link to the list of ontologies}}. This list comprehensively details each ontology's name, abbreviated short name, domain, projects utilizing it, purpose, publication of competency questions (CQs), licensing, last update date, homepage, ontology category, and a link to the ontology file. Additionally, it includes references to academic papers, citation counts, practical use cases, distinguishing features that contribute to its common use, and any special problems or challenges. 

Table~\ref{tab:ontologies_domains} highlights the distribution of ontologies across various MSE sub-domains, with Materials Representation and Materials Characterization having the most extensive coverage due to the complexity of these fields. Process Modeling, Nanomaterials, Computational Materials Science, and Materials Data also show significant ontology usage. However, domains such as Batteries, Chemistry, Energy Systems, Tribology, Biomaterials, and Sensors comprise fewer ontologies, indicating a need for further development to enhance their modeling capabilities and support advanced research applications.

\begin{table}[b]
\centering
\caption{List of ontologies in the domain of Materials Science and Engineering, categorized based on the subdomain.}
\label{tab:ontologies_domains}
\begin{adjustbox}{width=\textwidth}
\begin{tabular}{m{3cm}|m{2cm}|p{10cm}}
\hline
\textbf{Domain} & \textbf{Number of Ontologies} & \textbf{Ontologies} \\
\hline
Materials Representation & 14 & Material Science and Engineering Ontology, MatOnto, Standards-Specific Ontology Standard, Semantic Materials Manufacturing Design, Materials And Molecules Basic Ontology, Periodic Table Ontology, Metal Alloy, Building Material Ontology, Material properties ontology, Materials Ontology, Materials Vocabulary, Materials Data Science Ontology, PRovenance Information in MAterials science, OIE materials \\
\hline
Materials Characterization & 13 & EMMO Crystallography, CIF Core Ontology, Dislocation Ontology, Characterisation Methodology Domain Ontology (CHAMEO), EMMO Microstructure, Material Science Lab Equipment Ontology, Chemical Methods Ontology, Crystal Structure Ontology, Point Defects Ontology, Crystallographic Defect Core Ontology, Line Defect Ontology, Planar Defects Ontology, OIE Characterisation Methods \\
\hline
Process Modeling & 10 & EMMO General Process Ontology, PMD Core Ontology, SMART-Protocols, Baden Württemberg Material Digital Domain Mid Level Ontology, Baden Württemberg Material Digital Domain Ontology, Ontology of Scientific Experiments, Metadata4Ing Ontology, The Open Provenance Model for Workflows, Ontology to describe Workflows in Linked Data, Ontology for Simulation Modelling Optimization \\
\hline
Nanomaterials & 4 & NanoParticle Ontology, eNanoMapper, MaterialsMine, NanoMine \\
\hline
Computational Materials Science & 6 & EMMO Atomistic and Electronic Modelling, Materials Design Ontology, Computational Material Sample Ontology, Dislocation Simulation and Model Ontology, Atomistic Simulation Methods Ontology, OIE models \\
\hline
Materials Data & 5 & EMMO Datamodel ontology, EMMO Mappings ontology, MatWerk Ontology, Virtual Materials Marketplace (VIMMP) Ontology, OIE software \\
\hline
Mechanical Testing & 3 & EMMO Mechanical Testing, Matolab Tensile Test Ontology, MatoLab Brinell Test Ontology \\
\hline
Additive Manufacturing & 3 & Additive Manufacturing Ontology, Laser Powder Bed Fusion Ontology, OIE manufacturing \\
\hline
Batteries & 2 & EMMO Battery Interface Ontology, EMMO Battery Value Chain Ontology Ontology \\
\hline
Chemistry & 2 & Chemical Entities of Biological Interest, Chemical Information Ontology \\
\hline
Energy Systems & 1 & Smart Applications REFerence \\
\hline
Tribology & 1 & tribAIn Ontology \\
\hline
Biomaterials & 1 & The Devices, Experimental Scaffolds and Biomaterials Ontology \\
\hline
Sensors & 1 & Semantic Sensor Network Ontology \\
\hline
\end{tabular}
\end{adjustbox}
\end{table}
Among the ontologies identified, the top-level ontologies include: Basic Formal Ontology (BFO) \cite{bfo}, Elementary Multiperspective Material Ontology (EMMO) \cite{emmo}, the Descriptive Ontology for Linguistic and Cognitive Engineering (DOLCE) \cite{dolce}, and Suggested Upper Merged Ontology (SUMO) \cite{sumo}. The mid-level ontologies encompass PMD Core Ontology, Material Science and Engineering Ontology, Baden Württemberg Material Digital Domain Mid Level Ontology, EMMO Datamodel ontology, EMMO Mappings ontology, Materials Data Science Ontology, Ontology of Scientific Experiments, and The Open Provenance Model for Workflows. The application-level ontologies are MatoLab Brinell Test Ontology and Matolab Tensile Test Ontology.

Our analysis highlights the diversity and complexity of MSE ontologies, reflecting their varied applications and the necessity for rigorous evaluation methodologies. BFO is prominently reused in 16~ontologies, EMMO in 12, BWMD-MID in 3, NPO in 2, and SSN, PMD Core, and GPO in one each, indicating its broad applicability and acceptance in the community. This frequent reuse suggests that BFO and EMMO are considered high-quality foundational ontologies representing various aspects of MSE.

The study found that only nine of these ontologies explicitly published their competency questions: PRovenance Information in MAterials science, Crystal Structure Ontology, Computational Material Sample Ontology, Metadata4Ing Ontology, Open Energy Ontology, Materials Design Ontology, Dislocation Ontology, SMART-Protocols, and PMD Core Ontology. Publishing CQs is crucial for clarifying the ontology's intent, facilitating adaptability assessment, and ensuring human-readable answers \cite{wilson2022conceptual}. While formulating CQs is an integral part of ontology development, they are often not published along with ontologies, hindering evaluation efforts. Although modern design methodologies advocate for user stories, personas, and contextual statements beyond CQs \cite{presutti2012pattern}, none of the examined ontologies adopted these approaches. This emphasizes the need for a more comprehensive approach to capturing and addressing user needs and requirements.

\subsection{Evaluation of the MSE Ontologies}
\label{sec:evaluation_results_section}

The evaluation of MSE ontologies reveals varying levels of complexity and detail across different sub-domains, as shown in Tables~\ref{tab:base-metrics-1} and \ref{tab:base-metrics-2}. In Materials Characterization, ontologies like EMMO Crystallography and CIF-core demonstrate moderate axiom counts and high annotation axiom counts, indicating a balance between detailed representation and descriptive metadata. CHAMEO, with substantial object properties and annotation axioms, emphasizes detailed relationships and descriptive details. In Process Modeling, GPO, EXPO, and PMDCO have high axiom and class counts, reflecting their capability to model complex processes comprehensively. The high DL expressivity of GPO ($\mathcal{SROIQ(D)}$) indicates its advanced logical constructs and reasoning capabilities.

In Computational Materials Science, CMSO and MDO show diverse modeling approaches with substantial axiom counts, indicating detailed and comprehensive domain representation. Materials Representation ontologies like MatOnto and MSEO, with high axiom and class counts, suggest rich and detailed modeling capabilities across various sub-domains. For Nanomaterials, NPO stands out with extensive axioms and classes, indicative of detailed nanoparticle interaction modeling. eNanoMapper and NanoMine, with lower axiom counts and simpler DL expressivity, focus on specific nanomaterial aspects. In Mechanical Testing, MOL Brinell has a high axiom count but simpler DL expressivity ($\mathcal{AL(D)}$), suggesting a thorough yet straightforward representation. Additive Manufacturing ontologies, such as LPBFO, exhibit detailed and complex representations, whereas AMONTOLOGY offers a broader but simpler structure. In the Batteries domain, EMMO BattINFO and EMMO BVC present similar metrics, indicating detailed modeling capabilities.

Schema metrics provide insights into the richness and interconnectedness of these ontologies, as seen in Tables~\ref{tab:schema_metrics-1} and \ref{tab:schema_metrics-2}. High ACR and RR values denote detailed and interconnected models, which are essential for capturing the complexity of material characterization. However, these also bring increased computational challenges. High IR values suggest well-structured hierarchies, enhancing knowledge organization but potentially complicating ontology maintenance. The diversity in metrics indicates that while some ontologies are well-suited for detailed and complex modeling, others may offer more streamlined and efficient structures, highlighting the need for balance depending on specific application requirements. For instance, in the Materials Characterization domain, ontologies like CSO and PLDO show high Attribute Richness (AR) and Axiom Class Ratio (ACR), indicating detailed knowledge representation and logical definitions, making them suitable for applications requiring extensive detail. EMMO Crystallography and CSO excel in Inheritance Richness (IR), suggesting a well-structured hierarchical categorization, are beneficial for understanding domain taxonomy. CHAMEO and DISO are notable for their high Relationship Richness (RR), implying a well-connected structure that enhances relationship understanding. The choice of ontology depends on the application needs, with CSO, PLDO, and CDCO recommended for detailed logical reasoning, EMMO Crystallography and CSO for hierarchical structuring, and CHAMEO and DISO for comprehensive relationship mapping.

Graph metrics, detailed in Tables~\ref{tab:topology-metrics-1} and \ref{tab:topology-metrics-2}, provide a deeper understanding of the structural properties of the ontologies. Higher root cardinality (NoR) indicates a diversified foundational structure, enhancing the ontology's ability to cover a broad range of concepts. Conversely, a higher number of leaf classes (NoL) suggests greater granularity and specificity, crucial for capturing detailed knowledge within the domain. The number of External Classes (NoC) is calculated by comparing the ontology’s namespace with the namespaces of the referenced classes. Classes that have a different namespace than those present in the ontology are considered external. For ontologies that consist of multiple modules, the modules are considered part of the core ontology. Therefore, classes from these modules are not treated as external. To be classified as external, a class must have a namespace that differs from any of the namespaces used within the core ontology and its modules. This ensures that only truly external references are counted, reflecting the ontology's degree of interdependence with other distinct ontologies. The number of external classes (NoC) highlights the degree of interoperability and alignment with external standards, with a higher count indicating better integration but potentially adding complexity. Together, these metrics reveal the balance between foundational diversity, detail specificity, and interconnectivity, impacting the ontology's usability and effectiveness in representing complex domains. For instance, in the Materials Characterization domain, ontologies like EMMO Crystallography and MSLE exhibit high root cardinality, which may suggest a broad foundational structure. However, ontologies like CSO and CDCO have fewer root classes, potentially indicating a narrower scope. It is important to note that this metric is just a hint; a thorough examination of the ontologies' content is necessary to draw valid conclusions. An ontology with fewer root classes might still have a broader scope at deeper levels of the hierarchy, and the reduced number of root classes could result from a higher-level top classification. Ontologies such as EMMO Microstructure and OIE Characterisation Methods show high leaf cardinality, reflecting a high level of detail and specificity. In terms of external class references, CHAMEO and EMMO Microstructure demonstrate better interoperability with higher counts, whereas PODO and PLDO have fewer external references, implying less reliance on external ontologies. Depth and breadth metrics indicate complex hierarchical structures in ontologies like EMMO Crystallography, while others, such as CSO and CDCO, display simpler, more focused structures. Overall, these metrics highlight the diversity in foundational breadth, detail granularity, and external interconnectivity across the Materials Characterization ontologies, impacting their comprehensiveness.

For detailed evaluations in other domains, please refer to Appendix~\ref{appendix_schema} and \ref{appendix_graph}. The appendix includes the same evaluations applied to ontologies in different domains such as Biomaterials, Sensors, and Energy, providing a comprehensive understanding of their complexities and structures.

\section{Conclusion and Outlook}
\label{sec:Conclusion_outlook}

In conclusion, this study provides a comprehensive evaluation of 94 ontologies within the field of Materials Science and Engineering (MSE), offering a detailed analysis based on quality-control metrics. The findings highlight both the strengths and weaknesses of the evaluated ontologies, emphasizing their structural complexities, domain-specific relevance, and reuse of existing ontological frameworks. However, the study also identifies critical areas for improvement, such as the limited adoption of competency questions and ontology design patterns, and the need for better documentation and user support to address common pitfalls and enhance overall quality and usability.

Future work will focus on several key areas to further advance ontology development in MSE. Efforts will be made to systematically identify and extract Ontology Design Patterns (ODPs) to enhance quality and reusability. Additionally, further research will aim to comprehensively evaluate the completeness, domain coverage, and accuracy of these ontologies by extracting relevant domain terms and concepts within the subdomain of materials science. Moreover, the assessment of FAIRness (Findability, Accessibility, Interoperability, and Reusability) of the ontologies will be incorporated into future studies.

\begin{acknowledgments}
The authors thank the German Federal Ministry of Education and Research (BMBF) for financial support of the project \href{https://www.materialdigital.de/}{Innovation-Platform MaterialDigital} through project funding FKZ no: 13XP5094F (FIZ).
This publication was written by the NFDI consortium \href{https://nfdi-matwerk.de/}{NFDI-MatWerk} in the context of the work of the association German National Research Data Infrastructure (NFDI) e.V.. NFDI is financed by the Federal Republic of Germany and the 16 federal states and funded by the Federal Ministry of Education and Research (BMBF) -- funding code M532701 / the Deutsche Forschungsgemeinschaft (DFG, German Research Foundation) -- project number 460247524. 

The authors would also like to acknowledge Sven Hertling for his valuable assistance in discussing the reviews.
\end{acknowledgments}

\bibliography{sample-1col}

\begin{thebibliography}{87}
\expandafter\ifx\csname natexlab\endcsname\relax\def\natexlab#1{#1}\fi
\providecommand{\url}[1]{\texttt{#1}}
\providecommand{\href}[2]{#2}
\providecommand{\path}[1]{#1}
\providecommand{\DOIprefix}{doi:}
\providecommand{\ArXivprefix}{arXiv:}
\providecommand{\URLprefix}{URL: }
\providecommand{\Pubmedprefix}{pmid:}
\providecommand{\doi}[1]{\href{http://dx.doi.org/#1}{\path{#1}}}
\providecommand{\Pubmed}[1]{\href{pmid:#1}{\path{#1}}}
\providecommand{\bibinfo}[2]{#2}
\ifx\xfnm\relax \def\xfnm[#1]{\unskip,\space#1}\fi
\bibitem[{Zhang et~al.(2015)Zhang, Zhao, and Wang}]{zhangoverview}
\bibinfo{author}{X.~Zhang}, \bibinfo{author}{C.~Zhao}, \bibinfo{author}{X.~Wang},
\newblock \bibinfo{title}{A survey on knowledge representation in materials science and engineering: An ontological perspective},
\newblock \bibinfo{journal}{Computers in Industry} \bibinfo{volume}{73} (\bibinfo{year}{2015}) \bibinfo{pages}{8--22}. \DOIprefix\doi{10.1016/j.compind.2015.07.005}.
\bibitem[{Bayerlein(2022)}]{bernd_overview_2022}
\bibinfo{author}{B.~e.~a. Bayerlein},
\newblock \bibinfo{title}{A perspective on digital knowledge representation in materials science and engineering},
\newblock \bibinfo{journal}{Advanced Engineering Materials} \bibinfo{volume}{24} (\bibinfo{year}{2022}) \bibinfo{pages}{2101176}. \DOIprefix\doi{10.1002/adem.202101176}.
\bibitem[{De~Baas et~al.(2023)De~Baas, Nostro, Friis, Ghedini, Goldbeck, Paponetti, Pozzi, Sarkar, Yang, Zaccarini, and Toti}]{Gerhard_domain_os}
\bibinfo{author}{A.~De~Baas}, \bibinfo{author}{P.~D. Nostro}, \bibinfo{author}{J.~Friis}, \bibinfo{author}{E.~Ghedini}, \bibinfo{author}{G.~Goldbeck}, \bibinfo{author}{I.~M. Paponetti}, \bibinfo{author}{A.~Pozzi}, \bibinfo{author}{A.~Sarkar}, \bibinfo{author}{L.~Yang}, \bibinfo{author}{F.~A. Zaccarini}, \bibinfo{author}{D.~Toti},
\newblock \bibinfo{title}{Review and alignment of domain-level ontologies for materials science},
\newblock \bibinfo{journal}{IEEE Access} \bibinfo{volume}{11} (\bibinfo{year}{2023}) \bibinfo{pages}{120372--120401}. \DOIprefix\doi{10.1109/ACCESS.2023.3327725}.
\bibitem[{Wilson et~al.(2022)Wilson, Goonetillake, Indika, and Ginige}]{wilson2022conceptual}
\bibinfo{author}{R.~Wilson}, \bibinfo{author}{J.~Goonetillake}, \bibinfo{author}{W.~Indika}, \bibinfo{author}{A.~Ginige},
\newblock \bibinfo{title}{A conceptual model for ontology quality assessment},
\newblock \bibinfo{journal}{Semantic Web}  (\bibinfo{year}{2022}) \bibinfo{pages}{1--47}.
\bibitem[{Jackson(2019)}]{jackson2019robot}
\bibinfo{author}{R.~C. e.~a. Jackson},
\newblock \bibinfo{title}{Robot: a tool for automating ontology workflows},
\newblock \bibinfo{journal}{BMC bioinformatics} \bibinfo{volume}{20} (\bibinfo{year}{2019}) \bibinfo{pages}{1--10}.
\bibitem[{Lantow(2016)}]{lantow2016ontometrics}
\bibinfo{author}{B.~Lantow},
\newblock \bibinfo{title}{Ontometrics: Application of on-line ontology metric calculation},
\newblock in: \bibinfo{booktitle}{BIR Workshops}, volume \bibinfo{volume}{1684}, \bibinfo{year}{2016}, pp. \bibinfo{pages}{1--12}.
\bibitem[{Poveda-Villal{\'o}n(2014)}]{poveda2014oops}
\bibinfo{author}{M.~e.~a. Poveda-Villal{\'o}n},
\newblock \bibinfo{title}{Oops!(ontology pitfall scanner!): An on-line tool for ontology evaluation},
\newblock \bibinfo{journal}{International Journal on Semantic Web and Information Systems} \bibinfo{volume}{10} (\bibinfo{year}{2014}) \bibinfo{pages}{7--34}.
\bibitem[{Arp et~al.(2015)Arp, Smith, and Spear}]{bfo}
\bibinfo{author}{R.~Arp}, \bibinfo{author}{B.~Smith}, \bibinfo{author}{A.~D. Spear}, \bibinfo{title}{{Building Ontologies with Basic Formal Ontology}}, \bibinfo{publisher}{The MIT Press}, \bibinfo{year}{2015}. \DOIprefix\doi{10.7551/mitpress/9780262527811.001.0001}.
\bibitem[{Adamovic(2017)}]{emmo}
\bibinfo{author}{N.~e.~a. Adamovic},
\newblock \bibinfo{title}{Proceedings of the 4th world congress on integrated computational materials engineering (icme 2017)},
\newblock in: \bibinfo{booktitle}{4th World Congress on Integrated Computational Materials Engineering (ICME 2017)}, \bibinfo{organization}{Springer}, \bibinfo{year}{2017}.
\bibitem[{Pease et~al.(2002)Pease, Niles, and Li}]{sumo}
\bibinfo{author}{A.~Pease}, \bibinfo{author}{I.~Niles}, \bibinfo{author}{J.~Li},
\newblock \bibinfo{title}{The suggested upper merged ontology: A large ontology for the semantic web and its applic ations},
\newblock \bibinfo{year}{2002}. \URLprefix \url{https://api.semanticscholar.org/CorpusID:8595184}.
\bibitem[{Gangemi(2002)}]{dolce}
\bibinfo{author}{A.~e.~a. Gangemi},
\newblock \bibinfo{title}{Sweetening ontologies with dolce},
\newblock in: \bibinfo{editor}{A.~G{\'o}mez-P{\'e}rez}, \bibinfo{editor}{V.~R. Benjamins} (Eds.), \bibinfo{booktitle}{Knowledge Engineering and Knowledge Management: Ontologies and the Semantic Web}, \bibinfo{publisher}{Springer Berlin Heidelberg}, \bibinfo{year}{2002}, pp. \bibinfo{pages}{166--181}.
\bibitem[{Alam(2021)}]{pmdco}
\bibinfo{author}{M.~e.~a. Alam},
\newblock \bibinfo{title}{Ontology modelling for materials science experiments},
\newblock in: \bibinfo{booktitle}{Poster\&Demo track and Workshop on Ontology-Driven Conceptual Modelling of Digital Twins, Semantics 2021}, volume \bibinfo{volume}{2941}, \bibinfo{organization}{RWTH Aachen}, \bibinfo{year}{2021}, p.~\bibinfo{pages}{11}.
\bibitem[{Lebo et~al.(2013)Lebo, Sahoo, and McGuinness}]{prov-o}
\bibinfo{author}{T.~Lebo}, \bibinfo{author}{S.~Sahoo}, \bibinfo{author}{D.~McGuinness}, \bibinfo{title}{Prov-o: The {PROV} ontology}, \bibinfo{year}{2013}. \URLprefix \url{https://www.w3.org/TR/prov-o/}.
\bibitem[{Chen(2022)}]{mseo}
\bibinfo{author}{Y.~e.~a. Chen},
\newblock \bibinfo{title}{Ontopanel: A tool for domain experts facilitating visual ontology development and mapping for fair data sharing in materials testing},
\newblock \bibinfo{journal}{Integrating Materials and Manufacturing Innovation} \bibinfo{volume}{11} (\bibinfo{year}{2022}) \bibinfo{pages}{545--556}.
\bibitem[{Moreno~Torres(2021)}]{bwmd_review}
\bibinfo{author}{B.~e.~a. Moreno~Torres},
\newblock \bibinfo{title}{An ontology-based approach to enable data-driven research in the field of ndt in civil engineering},
\newblock \bibinfo{journal}{Remote Sensing} \bibinfo{volume}{13} (\bibinfo{year}{2021}). \DOIprefix\doi{10.3390/rs13122426}.
\bibitem[{Hagelien(2021)}]{emmo_datamodel}
\bibinfo{author}{T.~e.~a. Hagelien},
\newblock \bibinfo{title}{A practical approach to ontology-based data modelling for semantic interoperability},
\newblock in: \bibinfo{booktitle}{14th World Congress on Computational Mechanics}, \bibinfo{year}{2021}.
\bibitem[{French(2024)}]{mds-o}
\bibinfo{author}{R.~H. French}, \bibinfo{title}{Materials data science ontology}, \bibinfo{year}{2024}. \URLprefix \url{https://matportal.org/ontologies/MDS}.
\bibitem[{Soldatova and King(2006)}]{expo}
\bibinfo{author}{L.~N. Soldatova}, \bibinfo{author}{R.~D. King},
\newblock \bibinfo{title}{An ontology of scientific experiments},
\newblock \bibinfo{journal}{Journal of the royal society interface} \bibinfo{volume}{3} (\bibinfo{year}{2006}) \bibinfo{pages}{795--803}.
\bibitem[{Garijo and Gil(2011)}]{OPMW}
\bibinfo{author}{D.~Garijo}, \bibinfo{author}{Y.~Gil},
\newblock \bibinfo{title}{A new approach for publishing workflows: abstractions, standards, and linked data},
\newblock in: \bibinfo{booktitle}{Proceedings of the 6th workshop on Workflows in support of large-scale science}, \bibinfo{year}{2011}, pp. \bibinfo{pages}{47--56}.
\bibitem[{Garijo and Corcho(2013)}]{p-plan}
\bibinfo{author}{D.~Garijo}, \bibinfo{author}{O.~Corcho}, \bibinfo{title}{P-plan: A vocabulary for describing the plans of scientific processes}, \bibinfo{year}{2013}. \URLprefix \url{http://vocab.linkeddata.es/p-plan/index.html}.
\bibitem[{Thomas~Hagelien(2021)}]{emmo-mapping-o}
\bibinfo{author}{T.~R.~J. Thomas~Hagelien, Jesper~Friis}, \bibinfo{title}{Emmo mappings ontology}, \bibinfo{year}{2021}. \URLprefix \url{https://github.com/emmo-repo/domain-mappings}.
\bibitem[{Science and group(2024)}]{PRIMA}
\bibinfo{author}{M.~D. Science}, \bibinfo{author}{I.~group}, \bibinfo{title}{Provenance information in materials science (prima)}, \bibinfo{year}{2024}. \bibinfo{note}{\url{https://github.com/Materials-Data-Science-and-Informatics/MDMC-NEP-top-level-ontology}}.
\bibitem[{Hodgson et~al.(2018)Hodgson, Keller, and Markovic}]{qudt}
\bibinfo{author}{R.~Hodgson}, \bibinfo{author}{P.~Keller}, \bibinfo{author}{M.~Markovic}, \bibinfo{title}{Qudt - quantities, units, dimensions and data types ontologies}, \bibinfo{year}{2018}. \URLprefix \url{https://github.com/qudt/qudt-public-repo}.
\bibitem[{Adham~Hashibon(2022)}]{emmo-mechanical-o}
\bibinfo{author}{J.~F.~M. Adham~Hashibon, Daniele~Toti}, \bibinfo{title}{Emmo mechanical testing}, \bibinfo{year}{2022}. \URLprefix \url{https://github.com/emmo-repo/domain-mechanical-testing}.
\bibitem[{Jesper~Friis(2020)}]{emmo-microstructure-o}
\bibinfo{author}{S.~G. Jesper~Friis}, \bibinfo{title}{Emmo microstructure}, \bibinfo{year}{2020}. \URLprefix \url{https://github.com/jesper-friis/emmo-microstructure}.
\bibitem[{Giraldo(2017)}]{sp_ontology}
\bibinfo{author}{O.~L. e.~a. Giraldo},
\newblock \bibinfo{title}{Using semantics for representing experimental protocols},
\newblock \bibinfo{journal}{Journal of Biomedical Semantics} \bibinfo{volume}{8} (\bibinfo{year}{2017}).
\bibitem[{Bryan~Miller(2014)}]{MatOnto-o}
\bibinfo{author}{R.~G. Bryan~Miller, Brendan~Heussler}, \bibinfo{title}{Matonto}, \bibinfo{year}{2014}. \URLprefix \url{https://github.com/inovexcorp/MatOnto-Ontologies}.
\bibitem[{et~al.(2011)}]{npo_ontology}
\bibinfo{author}{D.~G.~T. et~al.},
\newblock \bibinfo{title}{Nanoparticle ontology for cancer nanotechnology research},
\newblock \bibinfo{journal}{Journal of Biomedical Informatics} \bibinfo{volume}{44} (\bibinfo{year}{2011}) \bibinfo{pages}{59--74}. \DOIprefix\doi{10.1016/j.jbi.2010.03.001}.
\bibitem[{Daniele(2015)}]{saref_ontology}
\bibinfo{author}{L.~e.~a. Daniele},
\newblock \bibinfo{title}{Created in close interaction with the industry: the smart appliances reference (saref) ontology},
\newblock in: \bibinfo{booktitle}{Formal Ontologies Meet Industry: 7th International Workshop, FOMI 2015}, \bibinfo{organization}{Springer}, \bibinfo{year}{2015}, pp. \bibinfo{pages}{100--112}.
\bibitem[{et~al.(2012)}]{ssn_ontology}
\bibinfo{author}{M.~C. et~al.},
\newblock \bibinfo{title}{The ssn ontology of the w3c semantic sensor network incubator group},
\newblock \bibinfo{journal}{Journal of Web Semantics} \bibinfo{volume}{17} (\bibinfo{year}{2012}) \bibinfo{pages}{25--32}. \DOIprefix\doi{10.1016/j.websem.2012.05.003}.
\bibitem[{Del~Nostro et~al.(2022)Del~Nostro, Goldbeck, and Toti}]{CHAMEO}
\bibinfo{author}{P.~Del~Nostro}, \bibinfo{author}{G.~Goldbeck}, \bibinfo{author}{D.~Toti},
\newblock \bibinfo{title}{Chameo: An ontology for the harmonisation of materials characterisation methodologies},
\newblock \bibinfo{journal}{Applied Ontology}  (\bibinfo{year}{2022}) \bibinfo{pages}{1--21}.
\bibitem[{Zhao(2018)}]{nanomine_ontology}
\bibinfo{author}{H.~e.~a. Zhao},
\newblock \bibinfo{title}{Nanomine schema: An extensible data representation for polymer nanocomposites},
\newblock \bibinfo{journal}{APL Materials} \bibinfo{volume}{6} (\bibinfo{year}{2018}) \bibinfo{pages}{111108}. \DOIprefix\doi{10.1063/1.5046839}.
\bibitem[{GPO(2022)}]{GPO}
\bibinfo{title}{Emmo general process ontology}, \bibinfo{year}{2022}. \URLprefix \url{https://github.com/General-Process-Ontology/ontology}.
\bibitem[{BVC(2022)}]{BVC}
\bibinfo{title}{Battery value chain ontology}, \bibinfo{year}{2022}. \URLprefix \url{https://github.com/Battery-Value-Chain-Ontology/ontology}.
\bibitem[{EMM(2021)}]{EMMO-Crystallography}
\bibinfo{title}{Emmo crystallography}, \bibinfo{year}{2021}. \URLprefix \url{https://github.com/emmo-repo/domain-crystallography}.
\bibitem[{CIF(2022)}]{CIF-Core}
\bibinfo{title}{Cif core ontology}, \bibinfo{year}{2022}. \URLprefix \url{https://github.com/emmo-repo/CIF-ontology}.
\bibitem[{EMM(2021)}]{EMMO-Atomistic}
\bibinfo{title}{Emmo atomistic and electronic modelling}, \bibinfo{year}{2021}. \URLprefix \url{https://github.com/emmo-repo/domain-atomistic}.
\bibitem[{Ashino(2010)}]{matinfo}
\bibinfo{author}{T.~Ashino},
\newblock \bibinfo{title}{Materials ontology: An infrastructure for exchanging materials information and knowledge},
\newblock \bibinfo{journal}{Data Science Journal} \bibinfo{volume}{9} (\bibinfo{year}{2010}) \bibinfo{pages}{54--61}.
\bibitem[{SEM(2017)}]{SEMMD}
\bibinfo{title}{Semantic materials, manufacturing, and design}, \bibinfo{year}{2017}. \URLprefix \url{https://github.com/cpauloh/semmd}.
\bibitem[{Hastings(2015)}]{enanomapper}
\bibinfo{author}{J.~e.~a. Hastings},
\newblock \bibinfo{title}{enanomapper: harnessing ontologies to enable data integration for nanomaterial risk assessment},
\newblock \bibinfo{journal}{Journal of biomedical semantics} \bibinfo{volume}{6} (\bibinfo{year}{2015}) \bibinfo{pages}{1--15}.
\bibitem[{Ihsan(2021)}]{diso_ontology}
\bibinfo{author}{A.~Z. e.~a. Ihsan},
\newblock \bibinfo{title}{Steps towards a dislocation ontology for crystalline materials},
\newblock \bibinfo{journal}{arXiv preprint arXiv:2106.15136}  (\bibinfo{year}{2021}).
\bibitem[{Bat(2022)}]{BattINFO}
\bibinfo{title}{Emmo battery interface ontology}, \bibinfo{year}{2022}. \URLprefix \url{https://github.com/BIG-MAP/BattINFO}.
\bibitem[{A.~Guzmán(2024)}]{mwo}
\bibinfo{author}{S.~F. A.~Guzmán, A.Z.~Ihsan}, \bibinfo{title}{The matwerk ontology (mwo)}, \bibinfo{year}{2024}. \bibinfo{note}{\url{https://nfdi-matwerk.pages.rwth-aachen.de/ta-oms/mwo/docs/index.html}}.
\bibitem[{Piane(2021)}]{mambo_ontology}
\bibinfo{author}{F.~L. e.~a. Piane},
\newblock \bibinfo{title}{Introducing mambo: Materials and molecules basic ontology},
\newblock \bibinfo{journal}{arXiv preprint arXiv:2111.02482}  (\bibinfo{year}{2021}).
\bibitem[{Li(2022)}]{lpbf_ontology}
\bibinfo{author}{Z.~e.~a. Li},
\newblock \bibinfo{title}{A description logic based ontology for knowledge representation in process planning for laser powder bed fusion},
\newblock \bibinfo{journal}{Applied Sciences} \bibinfo{volume}{12} (\bibinfo{year}{2022}) \bibinfo{pages}{4612}.
\bibitem[{AMO(2021)}]{AMOntology}
\bibinfo{title}{Additive manufacturing ontology}, \bibinfo{year}{2021}. \URLprefix \url{https://github.com/iassouroko/AMontology}.
\bibitem[{Bui(2021)}]{Building-Material}
\bibinfo{title}{Building material ontology}, \bibinfo{year}{2021}. \URLprefix \url{https://matportal.org/ontologies/BUILDMAT}.
\bibitem[{Wheeler(2022)}]{ssos_ontology}
\bibinfo{author}{R.~L. Wheeler},
\newblock \bibinfo{title}{Niso ssos (standards-specific ontology standard): A new resource for standards publishers}  (\bibinfo{year}{2022}).
\bibitem[{Li et~al.(2020)Li, Armiento, and Lambrix}]{mdo}
\bibinfo{author}{H.~Li}, \bibinfo{author}{R.~Armiento}, \bibinfo{author}{P.~Lambrix},
\newblock \bibinfo{title}{An ontology for the materials design domain},
\newblock in: \bibinfo{booktitle}{The Semantic Web--ISWC 2020: 19th International Semantic Web Conference}, \bibinfo{organization}{Springer}, \bibinfo{year}{2020}, pp. \bibinfo{pages}{212--227}.
\bibitem[{Hakimi(2020)}]{deb_ontology}
\bibinfo{author}{O.~e.~a. Hakimi},
\newblock \bibinfo{title}{The devices, experimental scaffolds, and biomaterials ontology (deb): A tool for mapping, annotation, and analysis of biomaterials data},
\newblock \bibinfo{journal}{Advanced Functional Materials} \bibinfo{volume}{30} (\bibinfo{year}{2020}) \bibinfo{pages}{1909910}. \DOIprefix\doi{10.1002/adfm.201909910}.
\bibitem[{Horsch et~al.(2021)Horsch, Toti, Chiacchiera, Seaton, Goldbeck, and Todorov}]{OSMO}
\bibinfo{author}{M.~T. Horsch}, \bibinfo{author}{D.~Toti}, \bibinfo{author}{S.~Chiacchiera}, \bibinfo{author}{M.~A. Seaton}, \bibinfo{author}{G.~Goldbeck}, \bibinfo{author}{I.~T. Todorov},
\newblock \bibinfo{title}{Osmo: Ontology for simulation, modelling, and optimization}  (\bibinfo{year}{2021}).
\bibitem[{Mat(2022)}]{MatVoc}
\bibinfo{title}{Materials vocabulary}, \bibinfo{year}{2022}. \URLprefix \url{https://github.com/stream-project/ontology}.
\bibitem[{Booshehri(2021)}]{oeo_ontology}
\bibinfo{author}{M.~e.~a. Booshehri},
\newblock \bibinfo{title}{Introducing the open energy ontology: Enhancing data interpretation and interfacing in energy systems analysis},
\newblock \bibinfo{journal}{Energy and AI} \bibinfo{volume}{5} (\bibinfo{year}{2021}) \bibinfo{pages}{100074}.
\bibitem[{Ceusters and Smith(2015)}]{IAO}
\bibinfo{author}{W.~Ceusters}, \bibinfo{author}{B.~Smith},
\newblock \bibinfo{title}{Aboutness: Towards foundations for the information artifact ontology}  (\bibinfo{year}{2015}).
\bibitem[{Mungall(2020)}]{RO}
\bibinfo{author}{C.~e.~a. Mungall}, \bibinfo{title}{oborel/obo-relations: 2020-07-21}, \bibinfo{year}{2020}. \URLprefix \url{https://doi.org/10.5281/zenodo.3955125}. \DOIprefix\doi{10.5281/zenodo.3955125}.
\bibitem[{Gkoutos et~al.(2012)Gkoutos, Schofield, and Hoehndorf}]{UO}
\bibinfo{author}{G.~V. Gkoutos}, \bibinfo{author}{P.~N. Schofield}, \bibinfo{author}{R.~Hoehndorf},
\newblock \bibinfo{title}{The units ontology: a tool for integrating units of measurement in science},
\newblock \bibinfo{journal}{Database} \bibinfo{volume}{2012} (\bibinfo{year}{2012}) \bibinfo{pages}{bas033}.
\bibitem[{OMO(2014)}]{OMO-ontology}
\bibinfo{title}{Obo metadata ontology}, \bibinfo{year}{2014}. \URLprefix \url{https://github.com/information-artifact-ontology/ontology-metadata}.
\bibitem[{Kügler et~al.(2020)Kügler, Marian, Schleich, Tremmel, and Wartzack}]{tribAIn}
\bibinfo{author}{P.~Kügler}, \bibinfo{author}{M.~Marian}, \bibinfo{author}{B.~Schleich}, \bibinfo{author}{S.~Tremmel}, \bibinfo{author}{S.~Wartzack},
\newblock \bibinfo{title}{tribain—towards an explicit specification of shared tribological understanding},
\newblock \bibinfo{journal}{Applied Sciences} \bibinfo{volume}{10} (\bibinfo{year}{2020}). \URLprefix \url{https://www.mdpi.com/2076-3417/10/13/4421}. \DOIprefix\doi{10.3390/app10134421}.
\bibitem[{CMO(2023)}]{CMO}
\bibinfo{title}{Chemical methods ontology}, \bibinfo{year}{2023}. \URLprefix \url{https://github.com/rsc-ontologies/rsc-cmo}.
\bibitem[{CIO(2024)}]{CIO}
\bibinfo{title}{Chemical information ontology}, \bibinfo{year}{2024}. \URLprefix \url{https://github.com/egonw/semanticchemistry}.
\bibitem[{Rubiera and de~Sainte~Marie(2011)}]{ONTORULE}
\bibinfo{author}{E.~Rubiera}, \bibinfo{author}{C.~de~Sainte~Marie},
\newblock \bibinfo{title}{Ontorule: From business knowledge to ontology-and rules-based applications},
\newblock in: \bibinfo{booktitle}{Proceedings of the 2011 Conference of the Center for Advanced Studies on Collaborative Research}, \bibinfo{year}{2011}, pp. \bibinfo{pages}{327--329}.
\bibitem[{Pho(2022)}]{Photovoltaics}
\bibinfo{title}{Photovoltaics ontology}, \bibinfo{year}{2022}. \URLprefix \url{https://github.com/emmo-repo/domain-photovoltaics}.
\bibitem[{MDS(2024)}]{MDS}
\bibinfo{title}{Materials data science ontology}, \bibinfo{year}{2024}. \URLprefix \url{https://matportal.org/ontologies/MDS}.
\bibitem[{DSI(2024)}]{DSIM}
\bibinfo{title}{Dislocation simulation and model ontology}, \bibinfo{year}{2024}. \URLprefix \url{https://github.com/OCDO/DSIM}.
\bibitem[{Ihsan et~al.(2021)Ihsan, Dessì, Alam, Sack, and Sandfeld}]{CSO}
\bibinfo{author}{A.~Z. Ihsan}, \bibinfo{author}{D.~Dessì}, \bibinfo{author}{M.~Alam}, \bibinfo{author}{H.~Sack}, \bibinfo{author}{S.~Sandfeld}, \bibinfo{title}{Steps towards a dislocation ontology for crystalline materials}, \bibinfo{year}{2021}. \href{http://arxiv.org/abs/2106.15136}{{\tt arXiv:2106.15136}}.
\bibitem[{ASM(2024)}]{ASMO}
\bibinfo{title}{Atomistic simulation methods ontology}, \bibinfo{year}{2024}. \URLprefix \url{https://github.com/OCDO/asmo}.
\bibitem[{POD(2024)}]{PODO}
\bibinfo{title}{Point defects ontology}, \bibinfo{year}{2024}. \URLprefix \url{https://github.com/OCDO/podo}.
\bibitem[{CDC(2024)}]{CDCO}
\bibinfo{title}{Crystallographic defect core ontology}, \bibinfo{year}{2024}. \URLprefix \url{https://github.com/OCDO/cdco}.
\bibitem[{LDO(2024)}]{LDO}
\bibinfo{title}{Line defect ontology}, \bibinfo{year}{2024}. \URLprefix \url{https://github.com/OCDO/ldo}.
\bibitem[{PLD(2024)}]{PLDO}
\bibinfo{title}{Planar defects ontology}, \bibinfo{year}{2024}. \URLprefix \url{https://github.com/OCDO/pldo}.
\bibitem[{Jalali et~al.(2023)Jalali, Mail, Aversa, and Kübel}]{MSLE}
\bibinfo{author}{M.~Jalali}, \bibinfo{author}{M.~Mail}, \bibinfo{author}{R.~Aversa}, \bibinfo{author}{C.~Kübel},
\newblock \bibinfo{title}{Msle: An ontology for materials science laboratory equipment – large-scale devices for materials characterization},
\newblock \bibinfo{journal}{Materials Today Communications} \bibinfo{volume}{35} (\bibinfo{year}{2023}) \bibinfo{pages}{105532}. \URLprefix \url{https://www.sciencedirect.com/science/article/pii/S2352492823002222}. \DOIprefix\doi{https://doi.org/10.1016/j.mtcomm.2023.105532}.
\bibitem[{OIE(2022)}]{OIE-o}
\bibinfo{title}{Open innovation environment domain ontologies, characterisation methods}, \bibinfo{year}{2022}. \URLprefix \url{https://github.com/emmo-repo/OIE-Ontologies/}.
\bibitem[{Arndt et~al.(2022)Arndt, Farnbacher, Fuhrmans, Hachinger, Hickmann, Hoppe, Horsch, Iglezakis, Karmacharya, Lanza, Leimer, Munke, Terzijska, Theissen-Lipp, Wiljes, and Windeck}]{M4I}
\bibinfo{author}{S.~Arndt}, \bibinfo{author}{B.~Farnbacher}, \bibinfo{author}{M.~Fuhrmans}, \bibinfo{author}{S.~Hachinger}, \bibinfo{author}{J.~Hickmann}, \bibinfo{author}{N.~Hoppe}, \bibinfo{author}{M.~T. Horsch}, \bibinfo{author}{D.~Iglezakis}, \bibinfo{author}{A.~Karmacharya}, \bibinfo{author}{G.~Lanza}, \bibinfo{author}{S.~Leimer}, \bibinfo{author}{J.~Munke}, \bibinfo{author}{D.~Terzijska}, \bibinfo{author}{J.~Theissen-Lipp}, \bibinfo{author}{C.~Wiljes}, \bibinfo{author}{J.~Windeck}, \bibinfo{title}{{Metadata4Ing: An ontology for describing the generation of research data within a scientific activity.}}, \bibinfo{year}{2022}. \URLprefix \url{https://doi.org/10.5281/zenodo.5957104}. \DOIprefix\doi{10.5281/zenodo.5957104}.
\bibitem[{Mat(2021)}]{MaterialsMine}
\bibinfo{title}{Materialsmine ontology}, \bibinfo{year}{2021}. \URLprefix \url{https://github.com/tetherless-world/materialsmine}.
\bibitem[{Per(2004)}]{Periodictable}
\bibinfo{title}{Periodic table ontology}, \bibinfo{year}{2004}. \URLprefix \url{http://www.daml.org/2003/01/periodictable/}.
\bibitem[{K{\"a}fer et~al.(2018)K{\"a}fer, Lauber, and Harth}]{WILD}
\bibinfo{author}{T.~K{\"a}fer}, \bibinfo{author}{S.~Lauber}, \bibinfo{author}{A.~Harth},
\newblock \bibinfo{title}{Using workflows to build compositions of read-write linked data apis on the web of things.},
\newblock in: \bibinfo{booktitle}{ISWC (P\&D/Industry/BlueSky)}, \bibinfo{year}{2018}.
\bibitem[{Bandrowski et~al.(2016)Bandrowski, Brinkman, Brochhausen, Brush, Bug, Chibucos, Clancy, Courtot, Derom, Dumontier et~al.}]{OBI}
\bibinfo{author}{A.~Bandrowski}, \bibinfo{author}{R.~Brinkman}, \bibinfo{author}{M.~Brochhausen}, \bibinfo{author}{M.~H. Brush}, \bibinfo{author}{B.~Bug}, \bibinfo{author}{M.~C. Chibucos}, \bibinfo{author}{K.~Clancy}, \bibinfo{author}{M.~Courtot}, \bibinfo{author}{D.~Derom}, \bibinfo{author}{M.~Dumontier}, et~al.,
\newblock \bibinfo{title}{The ontology for biomedical investigations},
\newblock \bibinfo{journal}{PloS one} \bibinfo{volume}{11} (\bibinfo{year}{2016}) \bibinfo{pages}{e0154556}.
\bibitem[{Ashburner et~al.(2000)Ashburner, Ball, Blake, Botstein, Butler, Cherry, Davis, Dolinski, Dwight, Eppig et~al.}]{GO}
\bibinfo{author}{M.~Ashburner}, \bibinfo{author}{C.~A. Ball}, \bibinfo{author}{J.~A. Blake}, \bibinfo{author}{D.~Botstein}, \bibinfo{author}{H.~Butler}, \bibinfo{author}{J.~M. Cherry}, \bibinfo{author}{A.~P. Davis}, \bibinfo{author}{K.~Dolinski}, \bibinfo{author}{S.~S. Dwight}, \bibinfo{author}{J.~T. Eppig}, et~al.,
\newblock \bibinfo{title}{Gene ontology: tool for the unification of biology},
\newblock \bibinfo{journal}{Nature genetics} \bibinfo{volume}{25} (\bibinfo{year}{2000}) \bibinfo{pages}{25--29}.
\bibitem[{Gkoutos et~al.(2018)Gkoutos, Schofield, and Hoehndorf}]{PATO}
\bibinfo{author}{G.~V. Gkoutos}, \bibinfo{author}{P.~N. Schofield}, \bibinfo{author}{R.~Hoehndorf},
\newblock \bibinfo{title}{The anatomy of phenotype ontologies: principles, properties and applications},
\newblock \bibinfo{journal}{Briefings in Bioinformatics} \bibinfo{volume}{19} (\bibinfo{year}{2018}) \bibinfo{pages}{1008--1021}.
\bibitem[{Mat(2020)}]{Material-Properties}
\bibinfo{title}{Material properties ontology}, \bibinfo{year}{2020}. \URLprefix \url{https://bimerr.iot.linkeddata.es/def/material-properties/}.
\bibitem[{MOL(2021)}]{MOL-TENSILE}
\bibinfo{title}{Matolab tensile test ontology}, \bibinfo{year}{2021}. \URLprefix \url{https://matportal.org/ontologies/MOL_TENSILE}.
\bibitem[{MOL(2022)}]{MOL-BRINELL}
\bibinfo{title}{Matolab brinell test ontology}, \bibinfo{year}{2022}. \URLprefix \url{https://matportal.org/ontologies/MOL_BRINELL}.
\bibitem[{et~al.(2020)}]{vimmp_ontology}
\bibinfo{author}{M.~T.~H. et~al.}, \bibinfo{title}{Ontologies for the virtual materials marketplace}, \bibinfo{year}{2020}. \href{http://arxiv.org/abs/1912.01519}{{\tt arXiv:1912.01519}}.
\bibitem[{Degtyarenko et~al.(2007)Degtyarenko, De~Matos, Ennis, Hastings, Zbinden, McNaught, Alc{\'a}ntara, Darsow, Guedj, and Ashburner}]{chebi}
\bibinfo{author}{K.~Degtyarenko}, \bibinfo{author}{P.~De~Matos}, \bibinfo{author}{M.~Ennis}, \bibinfo{author}{J.~Hastings}, \bibinfo{author}{M.~Zbinden}, \bibinfo{author}{A.~McNaught}, \bibinfo{author}{R.~Alc{\'a}ntara}, \bibinfo{author}{M.~Darsow}, \bibinfo{author}{M.~Guedj}, \bibinfo{author}{M.~Ashburner},
\newblock \bibinfo{title}{Chebi: a database and ontology for chemical entities of biological interest},
\newblock \bibinfo{journal}{Nucleic acids research} \bibinfo{volume}{36} (\bibinfo{year}{2007}) \bibinfo{pages}{D344--D350}.
\bibitem[{CHE(2024)}]{CHEMINF}
\bibinfo{title}{Chemical information ontology}, \bibinfo{year}{2024}. \URLprefix \url{https://github.com/egonw/semanticchemistry}.
\bibitem[{Met(2021)}]{MetalAlloy}
\bibinfo{title}{Metal alloy}, \bibinfo{year}{2021}. \URLprefix \url{https://github.com/arsarkar/ontorepo/tree/master/metal-alloy}.
\bibitem[{Presutti et~al.(2012)Presutti, Blomqvist, Daga, and Gangemi}]{presutti2012pattern}
\bibinfo{author}{V.~Presutti}, \bibinfo{author}{E.~Blomqvist}, \bibinfo{author}{E.~Daga}, \bibinfo{author}{A.~Gangemi},
\newblock \bibinfo{title}{Pattern-based ontology design},
\newblock \bibinfo{journal}{Ontology Engineering in a Networked World}  (\bibinfo{year}{2012}) \bibinfo{pages}{35--64}.

\end{thebibliography}
\newpage
\appendix
\section*{Supplementary Information}
This supplementary section provides a comprehensive evaluation of various aspects of MSE ontologies. Section~\ref{appendix_pitfalls} focuses on error detection using the OOPS! tool. Section~\ref{appendix_base} presents base metrics, examining structural aspects. Section~\ref{appendix_schema} evaluates schema metrics. Finally, Section~\ref{appendix_graph} details graph metrics.

\section{Error Detection}
\label{appendix_pitfalls}

The OOPS! (Ontology Pitfall Scanner!) tool was employed for error-checking evaluations. Identifying and categorizing these pitfalls is essential for the evaluation process. Critical pitfalls impact ontology consistency, reasoning, and applicability, leading to significant issues in how the ontology is used and interpreted. Important pitfalls, while not affecting the core functionality, can still degrade the quality and reliability of the ontology. Minor pitfalls are less problematic but addressing them can enhance the ontology's organization and user-friendliness. By comparing the presence of these pitfalls across different ontologies, MSE domain experts can make informed decisions about which ontology best meets their needs, ensuring that the chosen ontology is robust, reliable, and suitable for their specific applications. Table \ref{tab:pitfalls} describes only the pitfalls that exist in the MSE ontologies, along with their descriptions and impact levels.

Table \ref{tab:evaluation_oops} details the results of ontology evaluation using the OOPS! tool. Critical pitfalls affecting ontology consistency and reasoning include P19 (Multiple domains or ranges for properties), P40 (Namespace hijacking), P31 (Incorrect use of owl:equivalentClass), P05 (Incorrect use of owl:inverseOf), P29 (Incorrect use of owl:TransitiveProperty), and P27 (Incorrect use of owl:equivalentProperty). Important pitfalls, although less severe, still affect ontology quality and should be addressed. Minor pitfalls, while not critical, can be improved for better organization and user-friendliness.

The frequent occurrence of critical pitfalls such as P19 and P40 across multiple ontologies suggests a need for more stringent guidelines and validation tools during ontology development. The presence of these pitfalls can significantly impact the usability and accuracy of the ontologies in real-world applications. For instance, P19's misinterpretation issues can lead to incorrect inferences, while P40 can hinder effective data retrieval and integration.

\begin{table}[t]
\centering
\caption{Identified Pitfalls \cite{poveda2014oops} in the MSE Ontologies.}
\label{tab:pitfalls}
\begin{adjustbox}{width=\textwidth}
\begin{tabular}{>{\centering\arraybackslash}m{.6cm}|>{\centering\arraybackslash}m{1cm}|m{6cm}|m{3cm}}
\hline
\textbf{Imp.} & \textbf{Pitfall} & \textbf{Description} & \textbf{Impact} \\
\hline
\multirow{6}{*}{\rotatebox[origin=c]{90}{\parbox{6cm}{\centering Critical}}}
& P19 & Multiple domains or ranges defined for properties: Leads to misinterpretation as a conjunction in OWL. & Affects consistency and reasoning \\ \cline{2-4}
& P40 & Namespace hijacking: Terms from another namespace are used without proper definition, preventing information retrieval. & Hinders information retrieval \\ \cline{2-4}
& P31 & Incorrect use of owl:equivalentClass: Defining non-equivalent classes as equivalent. & Affects class definition accuracy \\ \cline{2-4}
& P05 & Incorrect use of owl:inverseOf: Defining non-inverse relationships as inverse. & Impacts relationship integrity \\ \cline{2-4}
& P29 & Incorrect use of owl:TransitiveProperty: Defining non-transitive relationships as transitive. & Distorts relationship hierarchy \\ \cline{2-4}
& P27 & Incorrect use of owl:equivalentProperty: Defining non-equivalent properties as equivalent. & Compromises property definitions \\ \hline

\multirow{3}{*}{\rotatebox[origin=c]{90}{\parbox{3cm}{\centering Important}}}
& P11 & Lack of domain or range definitions: Properties without defined domain or range may cause misunderstandings. & Leads to ambiguous property usage \\ \cline{2-4}
& P24 & Defining classes as instances: Instances incorrectly defined as classes, causing confusion in class hierarchy. & Confuses class hierarchy \\ \cline{2-4}
& P30 & Misuse of transitive property: Improper use of transitive property affects ontology reasoning. & Affects transitive reasoning \\ \hline

\multirow{3}{*}{\rotatebox[origin=c]{90}{\parbox{3cm}{\centering Minor}}}
& P04 & Misuse of symmetric property: Misuse can lead to incorrect inference. & Leads to incorrect inference \\ \cline{2-4}
& P08 & Using different naming conventions: Inconsistent naming conventions reduce ontology clarity. & Reduces clarity \\ \cline{2-4}
& P13 & Use of deprecated classes or properties: Use of deprecated elements affects ontology maintenance. & Affects maintenance \\ 
\hline
\end{tabular}
\end{adjustbox}
\end{table}

\begin{table}[t]
\centering
\caption{The results of error checking evaluation of ontologies using OOPS! tool. The numbers indicate the frequency of each type of pitfall (P) encountered in the ontologies.}
\label{tab:evaluation_oops}
\begin{adjustbox}{width=\textwidth}
\begin{tabular}{m{4cm}|ccc|cccc|cccc}
\hline
\multirow{2}{*}{\textbf{Ontology Name}} & \multicolumn{3}{c|}{\textbf{Critical}} & \multicolumn{4}{c|}{\textbf{Important}} & \multicolumn{4}{c}{\textbf{Minor}} \\ 
\cline{2-4} \cline{5-8} \cline{9-12}
 & \textbf{P19} & \textbf{P29} & \textbf{P05} & \textbf{P11} & \textbf{P24} & \textbf{P30} & \textbf{P34} & \textbf{P04} & \textbf{P08} & \textbf{P13} & \textbf{P36} \\ 
\hline
AMONTOLOGY & 9 & - & - & 2 & 1 & - & 60 & 2 & 305 & 9 & 1 \\ 
ASMO & - & - & - & 2 & - & 1 & - & 1 & 33 & 14 & 1 \\ 
BWMD-DOMAIN & 1 & - & - & 30 & - & 13 & - & 1 & 37 & 24 & 1 \\ 
BWMD-MID & 1 & - & - & 30 & - & 12 & - & 1 & 37 & 24 & 1 \\ 
CDCO & - & - & - & - & - & - & 1 & 3 & 3 & - & 1 \\ 
CHAMEO & - & - & - & - & - & - & - & - & - & - & 1 \\ 
CIF-core & - & - & - & 5 & - & 1 & - & 20 & 72 & 4 & 1 \\ 
CMSO & - & - & - & 3 & - & 1 & 1 & 1 & 73 & 13 & 1 \\ 
CSO & - & - & - & 4 & - & - & - & - & 31 & 19 & 1 \\ 
DEB & - & - & - & - & - & - & - & - & - & - & - \\ 
DISO & - & - & - & - & - & - & - & - & - & - & - \\ 
DSIM & - & - & - & - & - & - & - & - & - & - & - \\ 
EMMO Atomistic & - & - & - & - & - & - & - & - & - & - & - \\ 
EMMO Crystallography & - & - & - & - & - & - & - & - & - & - & - \\ 
EMMO Mappings & - & - & - & - & - & - & - & - & - & - & - \\ 
EMMO Mechanical Testing & - & - & - & - & - & - & - & - & - & - & - \\ 
EMMO Microstructure & - & - & - & - & - & - & - & - & - & - & - \\ 
EXPO & - & - & - & - & - & - & - & - & - & - & - \\ 
GPO & - & - & - & - & - & - & - & - & - & - & - \\ 
IAO & - & - & - & - & - & - & - & - & - & - & - \\ 
LDO & - & - & - & - & - & - & 1 & 3 & 6 & - & 1 \\ 
LPBFO & 1 & - & - & 31 & - & 13 & - & 1 & 46 & 25 & 1 \\ 
MAMBO & 2 & 3 & - & 7 & - & 1 & 1 & 5 & 153 & 26 & 1 \\ 
MDO & - & - & - & 8 & - & - & - & 2 & 15 & 32 & 1 \\ 
MDS & - & - & - & - & - & - & - & - & - & - & - \\ 
MOL BRINELL & - & - & - & 21 & - & - & - & 23 & 58 & 17 & 1 \\ 
MOL TENSILE & 1 & - & - & 91 & - & 13 & - & 1 & 116 & 81 & 1 \\ 
MSEO & - & - & - & - & - & - & - & - & - & - & - \\ 
MSLE & - & - & - & - & - & - & - & - & - & - & - \\ 
MaterialsMine & - & - & - & 3 & - & - & - & 2 & 20 & 25 & 1 \\ 
NanoMine & - & - & - & - & - & - & - & - & - & - & 1 \\ 
OA & - & - & - & - & - & - & 3 & - & - & - & 1 \\ 
OBO & - & - & 5 & 17 & - & 1 & 1 & - & - & 6 & 1 \\ 
OEO & - & - & - & - & - & - & - & - & - & - & - \\ 
OSMO & 5 & 2 & - & - & - & - & 20 & - & 318 & 119 & 1 \\ 
PLDO & - & - & - & 3 & - & 1 & - & 1 & 17 & 2 & 1 \\ 
PMDCO & - & - & - & - & - & - & - & - & - & - & - \\ 
PODO & - & - & - & 4 & - & 1 & 1 & 1 & 11 & - & 1 \\ 
PRIMA & - & - & - & 11 & - & - & - & 2 & 29 & 15 & 1 \\ 
QUDT & - & - & - & 217 & 3 & 1 & 10 & 1 & 139 & 82 & 1 \\ 
QUDV & - & - & - & 25 & - & 1 & 1 & 1 & 43 & 12 & 1 \\ 
SAREF & - & - & - & 12 & 2 & 1 & - & - & - & 22 & 1 \\ 
SKOS MDO & 1 & - & - & - & - & - & 1 & 2 & 2 & - & 1 \\ 
SP & 3 & - & - & 21 & 3 & 3 & - & 1 & 249 & 39 & 1 \\ 
SSN & - & - & - & 23 & 1 & 1 & - & 16 & 3 & 3 & 1 \\ 
WILD & - & - & - & - & - & - & 2 & - & - & - & 1 \\ 
\hline
\end{tabular}
\end{adjustbox}
\end{table}

\section{Base Metrics}
\label{appendix_base}

The Tables \ref{tab:base-metrics-1} and \ref{tab:base-metrics-2} provide an evaluation of the base metrics for MSE ontologies. It includes several columns such as Domain, Ontology Name, and the total number of Axioms. Additionally, it details the Class Count, Object Property (OP) Count, Data Property (DP) Count, and Annotation Axiom (Ann. Axm.) Count. The table also highlights the Description Logic Expressivity\footnote{ \url{https://en.wikipedia.org/wiki/Description_logic}} (DL Expr.) and includes information on the OWL2 profile\footnote{\url{https://www.w3.org/TR/owl2-profiles/}}. 

\begin{table}[t]
\centering
\caption{Base metrics evaluation of MSE ontologies (Part 1). The columns include Domain, Ontology Name, Axioms (total number of axioms), Class Count (total number of classes), Object Property (OP) Count, Data Property (DP) Count, Annotation Axiom (Ann. Axm.) Count, Description Logic Expressivity (DL Expr.), and OWL2 profile information (OWL2 P.).}
\label{tab:base-metrics-1}
\begin{adjustbox}{width=\textwidth}
\begin{tabular}{m{2cm}|m{4cm}>{\centering\arraybackslash}m{2cm}>{\centering\arraybackslash}m{1cm}>{\centering\arraybackslash}m{1cm}>{\centering\arraybackslash}m{0.9cm}>{\centering\arraybackslash}m{2cm}>{\centering\arraybackslash}m{2cm}>{\centering\arraybackslash}m{2cm}}
\hline
\textbf{Domain} & \textbf{Ontology Name} & \textbf{Axioms} & \textbf{Class} & \textbf{OP} & \textbf{DP} & \textbf{Ann. Axm.} & \textbf{DL Expr.} & \textbf{OWL2 P.} \\
\hline
\multirow{12}{2cm}{Materials Characterization}
& EMMO Crystallography & 357 & 61 & 5 & 1 & 175 & $\mathcal{ALCIQ}(\mathcal{D})$& OWL2 \\
& EMMO Microstructure & 183 & 61 & 2 & 0 & 60 & $\mathcal{ALE}$& OWL2 \\
& CIF-core & 321 & 31 & 1 & 1 & 176 & $\mathcal{AL(D)}$& OWL2-DL \\
& DISO & 373 & 38 & 33 & 12 & 147 & $\mathcal{ALCHIQ(D)}$& OWL2-DL \\
& CHAMEO & 491 & 74 & 44 & 2 & 234 & $\mathcal{ALCH(D)}$& OWL2 \\
& MSLE & 181 & 82 & 0 & 2 & 0 & $\mathcal{ALC}$& OWL2-DL \\
& CSO & 374 & 136 & 30 & 25 & 19 & $\mathcal{ALCIQ(D)}$& OWL2 \\
& PODO & 176 & 16 & 10 & 0 & 5 & $\mathcal{AL(D)}$& OWL2-RL \\
& CDCO & 60 & 3 & 0 & 0 & 34 & $\mathcal{AL}$& OWL2-RL \\
& LDO & 69 & 6 & 0 & 0 & 37 & $\mathcal{AL}$& OWL2-RL \\
& PLDO & 212 & 30 & 11 & 2 & 7 & $\mathcal{ALH(D)}$& OWL2 \\
& OIE Characterisation Methods & 129 & 44 & 0 & 0 & 44 & $\mathcal{AL}$& OWL2 \\ \hline

\multirow{10}{2cm}{Process Modeling} 
& GPO & 6249 & 963 & 86 & 5 & 2532 & $\mathcal{SROIQ(D)}$& OWL2 \\
& EXPO & 2067 & 325 & 78 & 0 & 646 & $\mathcal{ALCHN}$& OWL2-DL \\
& PMDCO & 2154 & 264 & 36 & 9 & 1454 & $\mathcal{ALCHIF(D)}$& OWL2 \\
& SMART-Protocols & 2999 & 399 & 43 & 0 & 1781 & $\mathcal{SHI}$& OWL2 \\
& BWMD-MID & 1546 & 336 & 27 & 12 & 771 & $\mathcal{ALCHI(D)}$& OWL2-DL \\
& BWMD-DOMAIN & 1800 & 459 & 0 & 0 & 917 & $\mathcal{AL}$& OWL2-DL \\
& OPMW & 202 & 96 & 21 & 22 & 18 & $\mathcal{ALUHF(D)}$& OWL2 \\
& WILD & 73 & 19 & 13 & 1 & 2 & $\mathcal{SH(D)}$& OWL2 \\
& OSMO & 1786 & 173 & 152 & 46 & 360 & $\mathcal{SROIN(D)}$& OWL2 \\
& M4I & 1203 & 38 & 57 & 37 & 751 & $\mathcal{SRIN(D)}$& OWL2-DL \\ \hline

\multirow{6}{2cm}{Computational Materials Science} & ASMO & 519 & 36 & 18 & 4 & 327 & $\mathcal{ALCHI(D)}$& OWL2 \\
& CMSO & 508 & 40 & 19 & 28 & 270 & $\mathcal{ALUHI(D)}$& OWL2 \\
& DSIM & 492 & 41 & 45 & 33 & 185 & $\mathcal{SHIQ(D)}$& OWL2 \\
& OIE Models & 353 & 114 & 5 & 0 & 130 & $\mathcal{ALEHI}$& OWL2 \\
& EMMO Atomistic & 64 & 18 & 3 & 1 & 34 & $\mathcal{ALEH(D)}$& OWL2 \\
& MDO & 574 & 38 & 32 & 32 & 268 & $\mathcal{ALCQ(D)}$ & OWL2 \\ \hline

\multirow{12}{2cm}{Materials Representation} & MSEO & 890 & 150 & 2 & 0 & 618 & $\mathcal{ALH}$ & OWL2 \\
& MatOnto & 5235 & 848 & 83 & 13 & 1841 & $\mathcal{SHOIQ(D)}$& OWL2 \\
& MATINFO & 549 & 140 & 13 & 8 & 202 & $\mathcal{ALQ(D)}$& OWL2 \\
& MatVoc & 154 & 28 & 12 & 3 & 75 & $\mathcal{ALI(D)}$& OWL2 \\
& SSOS & 244 & 27 & 19 & 21 & 79 & $\mathcal{SHI(D)}$& OWL2-RL \\
& MAMBO & 400 & 57 & 35 & 63 & 57 & $\mathcal{SHIQ(D)}$& OWL2-DL \\
& Periodictable & 1756 & 7 & 6 & 7 & 2 & $\mathcal{ALUON(D)}$& OWL2-DL \\
& BMO & 362 & 26 & 56 & 7 & 44 & $\mathcal{ALCRI(D)}$& OWL2 \\
& MAT & 549 & 140 & 13 & 8 & 202 & $\mathcal{ALQ(D)}$& OWL2-DL \\
& MDS & 1698 & 256 & 11 & 1 & 890 & $\mathcal{AL(D)}$& OWL2 \\
& PRIMA & 146 & 54 & 21 & 15 & 2 & $\mathcal{ALCH}$& OWL2-DL \\
& OIE materials & 460 & 119 & 1 & 0 & 204 & $\mathcal{ALC}$& OWL2 \\ \hline

\end{tabular}
\end{adjustbox}
\end{table}

\begin{table}[t]
\centering
\caption{Base metrics evaluation of MSE ontologies (Part 2). The columns include Domain, Ontology Name, Axioms (total number of axioms), Class Count (total number of classes), Object Property Count (OP Ct.), Data Property Count (DP Ct.), Annotation Axiom Count (Ann. Axm. Ct.), Description Logic Expressivity (DL Expr.), and OWL2 profile information (OWL2 P.).}
\label{tab:base-metrics-2}
\begin{adjustbox}{width=\textwidth}
\begin{tabular}{m{2cm}|m{4cm}>{\centering\arraybackslash}m{2cm}>{\centering\arraybackslash}m{1cm}>{\centering\arraybackslash}m{1cm}>{\centering\arraybackslash}m{0.9cm}>{\centering\arraybackslash}m{2cm}>{\centering\arraybackslash}m{2cm}>{\centering\arraybackslash}m{2cm}}
\hline
\textbf{Domain} & \textbf{Ontology Name} & \textbf{Axioms} & \textbf{Class Ct.} & \textbf{OP Ct.} & \textbf{DP Ct.} & \textbf{Ann. Axm. Ct.} & \textbf{DL Expr.} & \textbf{OWL2 P.} \\
\hline
\multirow{4}{2cm}{Nano\-materials}
& NPO & 28924 & 1906 & 65 & 22 & 11343 & $\mathcal{SHIN(D)}$& OWL2 \\
& eNanoMapper & 2809 & 772 & 2 & 0 & 1341 & $\mathcal{ALE}$& OWL2 \\
& NanoMine & 815 & 172 & 1 & 0 & 429 & $\mathcal{ALEO}$& OWL2 \\
& MaterialsMine & 2154 & 264 & 36 & 9 & 1454 & $\mathcal{ALCHIF(D)}$& OWL2 \\
\hline

\multirow{3}{2cm}{Mechanical Testing}
& EMMO Mechanical Testing & 1740 & 393 & 13 & 6 & 657 & $\mathcal{ALCHIQ(D)}$& OWL2 \\
& MOL Brinell & 16349 & 37 & 17 & 4 & 191 & $\mathcal{AL(D)}$& OWL2 \\
& MOL Tensile & 354 & 35 & 61 & 7 & 120 & $\mathcal{ALCHF(D)}$& OWL2 \\
\hline

\multirow{3}{2cm}{Additive Manufacturing}
& AMONTOLOGY & 130 & 85 & 5 & 0 & 3 & $\mathcal{ALE}$& OWL2 \\
& LPBFO & 663 & 179 & 2 & 0 & 346 & $\mathcal{ALCH}$& OWL2-DL \\
& OIE manufacturing & 763 & 228 & 5 & 0 & 304 & $\mathcal{ALEHI}$& OWL2 \\
\hline

\multirow{2}{2cm}{Batteries}
& BattINFO & 442 & 137 & 10 & 0 & 203 & $\mathcal{ALEI}$& OWL2 \\
& EMMO BVC & 568 & 182 & 11 & 0 & 131 & $\mathcal{ALCH}$& OWL2 \\
\hline

\multirow{1}{2cm}{Biomaterials}
& DEB & 2135 & 601 & 12 & 109 & 296 & $\mathcal{ALH(D)}$& OWL2-RL \\
\hline

\multirow{1}{2cm}{Sensor}
& SSN & 313 & 16 & 21 & 2 & 248 & $\mathcal{ALI(D)}$& OWL2-RL \\
\hline
\multirow{1}{2cm}{Energy}
& SAREF & 631 & 81 & 35 & 5 & 264 & $\mathcal{ALCIQ(D)}$& OWL2-DL \\
\hline

\end{tabular}
\end{adjustbox}
\end{table}

\section{Schema Metrics}
\label{appendix_schema}
Materials Characterization ontologies such as EMMO Crystallography (NoR: 33, NoL: 4) and MSLE (NoR: 63, NoL: 6) show broad foundational structures and detailed coverage, indicated by deep (Max Depth: 4) and broad (Max Breadth: 1760) hierarchies. CHAMEO and OIE Characterisation Methods have high leaf cardinality (NoL: 34-35) and demonstrate good interoperability with moderate tangledness (0.05-0.12). Process Modeling ontologies like GPO (NoR: 115, NoL: 8) and EXPO (NoL: 202) reflect broad foundational coverage and significant depth (Max Depth: 8-12) with moderate tangledness (0.10-0.11). PMDCO shows a balanced structure with extensive roots (NoR: 194) and low tangledness (0.02).

In Computational Materials Science, ASMO (NoR: 28, NoL: 3) and CMSO (NoR: 31, NoL: 2) have broad structures with no tangledness. DSIM (NoR: 17, NoL: 4) exhibits moderate depth and higher tangledness (0.24). Materials Representation ontologies like MatOnto (NoR: 848, NoL: 13) and MSEO (NoR: 100, NoL: 5) suggest deep, wide structures with no tangledness. Nanomaterials ontologies such as NPO (NoR: 65, NoL: 5) and NanoMine (NoR: 1, NoL: 5) show extensive depth (Max Depth: 14) and breadth (Max Breadth: 284-1593) with moderate to low tangledness (0.03-0.31). Mechanical Testing ontologies like EMMO Mechanical Testing (NoR: 13, NoL: 7) and MOL Brinell (NoR: 17, NoL: 3) have moderate depth (Max Depth: 3-7) and tangledness (0.21). Additive Manufacturing ontologies like AMONTOLOGY (NoR: 5, NoL: 3) and LPBFO (NoR: 2, NoL: 4) show moderate depth and breadth with low tangledness. Batteries ontologies like BattINFO (NoR: 10, NoL: 3) and EMMO BVC (NoR: 11, NoL: 4) suggest broad foundational coverage with high horizontal coverage (Max Breadth: 1203-1781) and low tangledness (0.01-0.05).

\begin{table}[t]
\centering
\caption{Schema metrics evaluation of MSE ontologies (Part 1). The columns include Domain, Ontology Name, Attribute Richness (AR), Inheritance Richness (IR), Relationship Richness (RR), Axiom Class Ratio (ACR), and Equivalence Ratio (ER).}
\label{tab:schema_metrics-1}
\begin{adjustbox}{width=\textwidth}
\begin{tabularx}{\textwidth}{m{3cm}|m{5cm} *{5}{>{\centering\arraybackslash}m{0.9cm}}}
\hline
\textbf{Domain} & \textbf{Ontology Name} & \textbf{AR} & \textbf{IR} & \textbf{RR} & \textbf{ACR} & \textbf{ER} \\
\hline
\multirow{12}{2cm}{Materials Characterization}
& EMMO Crystallography & 0.00 & 1.84 & 0.03 & 0.20 & 0.04 \\
& EMMO Microstructure & 0.01 & 1.26 & 0.22 & 0.84 & 0.13 \\
& CIF-core & 0.03 & 1.04 & 0.10 & 4.72 & 0.04 \\
& DISO & 0.32 & 1.63 & 0.39 & 9.82 & 0.00 \\
& CHAMEO & 0.03 & 0.80 & 0.44 & 6.64 & 0.00 \\
& MSLE & 0.04 & 1.54 & 0.25 & 1.76 & 0.00 \\
& CSO & 0.63 & 1.63 & 0.35 & 12.47 & 0.00 \\
& PODO & 0.50 & 1.00 & 0.00 & 17.60 & 0.00 \\
& CDCO & 0.00 & 0.00 & 0.00 & 20.00 & 0.00 \\
& LDO & 0.00 & 0.50 & 0.00 & 11.50 & 0.00 \\
& PLDO & 0.64 & 0.91 & 0.17 & 19.27 & 0.00 \\
& OIE Characterisation Methods & 0.00 & 0.95 & 0.02 & 2.93 & 0.02 \\
\hline
\multirow{10}{2cm}{Process Modeling}
& GPO & 0.00 & 1.12 & 0.24 & 1.77 & 0.13 \\
& EXPO & 0.00 & 1.33 & 0.51 & 6.36 & 0.00 \\
& PMDCO & 0.04 & 1.00 & 0.36 & 7.72 & 0.14 \\
& SP & 0.00 & 1.19 & 0.20 & 7.52 & 0.00 \\
& BWMD-MID & 0.03 & 1.00 & 0.11 & 4.26 & 0.00 \\
& BWMD-DOMAIN & 0.02 & 1.00 & 0.06 & 2.29 & 0.00 \\
& OPMW & 0.56 & 1.42 & 0.57 & 3.16 & 0.03 \\
& WILD & 0.05 & 0.47 & 0.65 & 3.84 & 0.00 \\
& OSMO & 0.27 & 1.63 & 0.39 & 10.32 & 0.10 \\
& M4I & 0.97 & 1.11 & 0.58 & 31.66 & 0.03 \\
\hline
\multirow{6}{2cm}{Computational Materials Science}
& ASMO & 0.11 & 0.67 & 0.45 & 14.42 & 0.03 \\
& CMSO & 0.70 & 0.55 & 0.46 & 12.70 & 0.00 \\
& DSIM & 0.80 & 1.41 & 0.47 & 12.00 & 0.00 \\
& OIE Models & 0.01 & 1.07 & 0.25 & 1.05 & 0.12 \\
& EMMO Atomistic & 0.01 & 1.66 & 0.12 & 0.12 & 0.13 \\
& MDO & 0.84 & 1.92 & 0.31 & 15.11 & 0.00 \\
\hline
\multirow{12}{2cm}{Materials Representation}
& MSEO & 0.01 & 1.40 & 0.35 & 3.72 & 0.15 \\
& MatOnto & 0.02 & 1.40 & 0.31 & 6.17 & 0.33 \\
& MATINFO & 0.11 & 0.99 & 0.05 & 0.10 & 0.00 \\
& MatVoc & 0.78 & 0.85 & 0.52 & 9.04 & 0.00 \\
& SSOS & 0.78 & 0.85 & 0.52 & 9.04 & 0.00 \\
& MAMBO & 1.11 & 0.75 & 0.56 & 11.09 & 0.00 \\
& Periodictable & 1.00 & 3.14 & 0.29 & 250.86 & 0.43 \\
& BMO & 0.18 & 0.80 & 0.63 & 2.05 & 0.08 \\
& MAT & 0.06 & 1.24 & 0.07 & 3.92 & 0.00 \\
& MDS & 0.00 & 0.86 & 0.05 & 6.63 & 0.00 \\
& PRIMA & 0.10 & 1.81 & 0.28 & 6.95 & 0.00 \\
& OIE materials & 0.00 & 1.08 & 0.19 & 2.53 & 0.09 \\
\hline
\end{tabularx}
\end{adjustbox}
\end{table}

\begin{table}[t]
\centering
\caption{Schema metrics evaluation of MSE ontologies (Part 2). The columns include Domain, Ontology Name, Attribute Richness (AR), Inheritance Richness (IR), Relationship Richness (RR), Axiom Class Ratio (ACR), and Equivalence Ratio (ER).}
\label{tab:schema_metrics-2}
\begin{adjustbox}{width=\textwidth}
\begin{tabularx}{\textwidth}{m{3cm}|m{5cm} *{5}{>{\centering\arraybackslash}m{0.9cm}}}
\hline
\textbf{Domain} & \textbf{Ontology Name} & \textbf{AR} & \textbf{IR} & \textbf{RR} & \textbf{ACR} & \textbf{ER} \\
\hline
\multirow{4}{2cm}{Nano\-materials}
& NPO & 0.01 & 1.46 & 0.82 & 15.18 & 0.21 \\
& eNanoMapper & 0.00 & 0.93 & 0.02 & 3.64 & 0.01 \\
& NanoMine & 0.00 & 1.25 & 0.14 & 0.47 & 0.03 \\
& MaterialsMine & 0.00 & 1.25 & 0.14 & 0.47 & 0.03 \\
\hline
\multirow{3}{2cm}{Mechanical Testing}
& EMMO Mechanical Testing & 0.01 & 1.60 & 0.09 & 2.07 & 0.09 \\
& MOL Brinell & 0.11 & 0.38 & 0.55 & 441.86 & 0.00 \\
& MOL Tensile & 0.04 & 1.00 & 0.21 & 0.91 & 0.00 \\
\hline
\multirow{3}{2cm}{Additive Manufacturing}
& AMONTOLOGY & 0.02 & 1.92 & 0.07 & 0.41 & 0.06 \\
& LPBFO & 0.02 & 1.00 & 0.09 & 1.30 & 0.00 \\
& OIE manufacturing & 0.00 & 1.04 & 0.15 & 1.76 & 0.06 \\
\hline
\multirow{2}{2cm}{Batteries}
& BattINFO & 0.00 & 1.57 & 0.05 & 0.24 & 0.03 \\
& EMMO BVC & 0.00 & 1.63 & 0.10 & 0.42 & 0.09 \\
\hline
\multirow{1}{2cm}{Biomaterials}
& DEB & 0.18 & 1.11 & 0.03 & 3.55 & 0.01 \\
\hline
\multirow{1}{2cm}{Sensor}
& SSN & 0.12 & 0.00 & 1.00 & 19.56 & 0.00 \\
\hline
\multirow{1}{2cm}{Energy}
& SAREF & 0.06 & 1.81 & 0.21 & 7.79 & 0.00 \\
\hline
\end{tabularx}
\end{adjustbox}
\end{table}

\section{Graph Metrics}
\label{appendix_graph}
Materials Characterization ontologies like EMMO Crystallography (NoR: 33, NoL: 4) and MSLE (NoR: 63, NoL: 6) show extensive foundational structures and detailed coverage, with deep (Max Depth: 4) and broad (Max Breadth: 1760) hierarchies. CHAMEO and OIE Characterisation Methods have high leaf cardinality (NoL: 34-35) and good interoperability with moderate tangledness (0.05-0.12). Process Modeling ontologies like GPO (NoR: 115, NoL: 8) and EXPO (NoL: 202) reflect broad foundational coverage and significant depth (Max Depth: 8-12) with moderate tangledness (0.10-0.11). PMDCO shows balanced structure with extensive roots (NoR: 194) and low tangledness (0.02).

In Computational Materials Science, ASMO (NoR: 28, NoL: 3) and CMSO (NoR: 31, NoL: 2) indicate broad structures with no tangledness. DSIM (NoR: 17, NoL: 4) has moderate depth and higher tangledness (0.24). Materials Representation ontologies like MatOnto (NoR: 848, NoL: 13) and MSEO (NoR: 100, NoL: 5) suggest deep, wide structures with no tangledness. Nanomaterials ontologies like NPO (NoR: 65, NoL: 5) and NanoMine (NoR: 1, NoL: 5) show extensive depth (Max Depth: 14) and breadth (Max Breadth: 284-1593) with moderate to low tangledness (0.03-0.31). Mechanical Testing ontologies like EMMO Mechanical Testing (NoR: 13, NoL: 7) and MOL Brinell (NoR: 17, NoL: 3) have moderate depth (Max Depth: 3-7) and tangledness (0.21). Additive Manufacturing ontologies like AMONTOLOGY (NoR: 5, NoL: 3) and LPBFO (NoR: 2, NoL: 4) show moderate depth and breadth with low tangledness. Batteries ontologies like BattINFO (NoR: 10, NoL: 3) and EMMO BVC (NoR: 11, NoL: 4) suggest broad foundational coverage with high horizontal coverage (Max Breadth: 1203-1781) and low tangledness (0.01-0.05).

\begin{table}[t]
\centering
\caption{Topology metrics of MSE ontologies (Part 1). The columns include Domain, Ontology Name, Absolute/Average/Maximum Depth (Abs./Avg./Max. Depth), Absolute/Average/Maximum Breadth (Abs./Avg./Max. Breadth), Tangledness (Tngld.), Number of Root Classes (NoR), Number of External Classes (NoC), and Number of Leaf Classes (NoL).}
\label{tab:topology-metrics-1}
\begin{adjustbox}{width=\textwidth}
\begin{tabularx}{\textwidth}{m{2cm}|m{3cm}>{\centering\arraybackslash}m{2.5cm}>{\centering\arraybackslash}m{2.5cm}>{\centering\arraybackslash}m{0.7cm}*{3}{>{\centering\arraybackslash}m{0.56cm}}}
\hline
\textbf{Domain} & \textbf{Ontology Name} & \textbf{Abs./Avg./Max. Depth} & \textbf{Abs./Avg./Max. Breadth} & \textbf{Tngld.} & \textbf{NoR} & \textbf{NoC} & \textbf{NoL} \\
\hline
\multirow{12}{2cm}{Materials Characterization}
& EMMO Crystallography & 1869/1.04/4 & 1803/112.69/1760 & 0.01 & 33 & 15 & 4 \\
& EMMO Microstructure & 322/1.37/4 & 235/7.58/176 & 0.05 & 0 & 35 & 35 \\
& CIF-core & 193/2.01/6 & 96/4.80/51 & 0.10 & 12 & 14 & 6 \\
& DISO & 22/1.69/4 & 13/3.25/9 & 0.03 & 10 & 9 & 4 \\
& CHAMEO & 145/1.88/3 & 77/4.05/24 & 0.12 & 52 & 24 & 5 \\
& MSLE & 368/3.44/6 & 107/5.10/26 & 0.02 & 63 & 5 & 6 \\
& CSO & 2/1.00/1 & 2/2.00/2 & 0.00 & 2 & 2 & 1 \\
& PODO & 25/2.27/3 & 11/2.75/6 & 0.10 & 7 & 1 & 3 \\
& CDCO & 3/1.00/1 & 3/3.00/3 & 0.00 & 3 & 3 & 1 \\
& LDO & 11/1.83/3 & 6/2.00/3 & 0.00 & 4 & 3 & 3 \\
& PLDO & 27/2.45/4 & 11/2.75/5 & 0.00 & 8 & 1 & 4 \\
& OIE Characterisation Methods & 163/3.70/5 &  44/4.00/10 & 0.00 & 0 & 34 & 34 \\
\hline
\multirow{10}{2cm}{Process Modeling}
& GPO & 968/2.47/8 & 392/4.61/218 & 0.11 & 115 & 38 & 8 \\
& EXPO & 2036/6.26/12 & 325/2.62/15 & 0.10 & 0 & 202 & 202 \\
& PMDCO & 1029/3.50/6 & 294/4.08/51 & 0.02 & 194 & 8 & 6 \\
& SP & 3167/7.94/14 & 399/2.77/24 & 0.08 & 256 & 1 & 14 \\
& BWMD-MID & 1297/3.57/7 & 363/5.76/88 & 0.00 & 274 & 9 & 7 \\
& BWMD-DOMAIN & 1640/2.08/6 & 787/5.70/363 & 0.00 & 322 & 35 & 6 \\
& OPMW & 94/1.29/3 & 73/5.21/56 & 0.09 & 9 & 13 & 3 \\
& WILD & 31/1.63/3 & 19/3.80/10 & 0.00 & 15 & 10 & 3 \\
& OSMO & 912/3.75/6 & 243/2.96/25 & 0.48 & 108 & 25 & 6 \\
& M4I & 73/1.92/4 & 38/2.71/13 & 0.16 & 25 & 13 & 4 \\
\hline
\multirow{6}{2cm}{Computational Materials Science}
& ASMO & 61/1.74/3 & 35/4.38/12 & 0.00 & 28 & 12 & 3 \\
& CMSO & 60/1.54/2 & 39/4.33/18 & 0.00 & 31 & 18 & 2 \\
& DSIM & 48/1.78/4 & 27/2.45/12 & 0.24 & 17 & 12 & 4 \\
& OIE Models & 697/2.08/6 & 335/9.57/230 & 0.01 & 80 & 9 & 6 \\
& EMMO Atomistic & 550/1.03/4 & 534/48.55/523 & 0.01 & 8 & 8 & 4 \\
& MDO & 92/2.42/4 & 38/4.75/24 & 0.47 & 31 & 1 & 4 \\
\hline
\multirow{12}{2cm}{Materials Representation}
& MSEO & 476/1.99/5 & 239/4.69/115 & 0.00 & 100 & 26 & 5 \\
& MatOnto & 1640/2.08/6 & 787/5.70/363 & 0.00 & 848 & 4 & 13 \\
& MATINFO & 609/1.00/2 & 606/303.00/603 & 0.00 & 140 & 7 & 2 \\
& MatVoc & 68/2.52/3 & 27/3.38/8 & 0.00 & 28 & 4 & 3 \\
& SSOS & 193/2.01/6 & 96/4.80/51 & 0.10 & 27 & 4 & 3 \\
& MAMBO & 94/1.77/3 & 53/3.31/17 & 0.04 & 57 & 17 & 3 \\
& Periodictable & 4/1.00/1 & 4/4.00/4 & 0.00 & 7 & 4 & 1 \\
& BMO & 206/1.16/3 & 177/35.40/157 & 0.00 & 26 & 6 & 3 \\
& MAT & 0/0.00/0 & 0/0.00/0 & 0.00 & 140 & 0 & 2 \\
& MDS & 677/2.64/4 & 256/4.34/36 & 0.00 & 256 & 36 & 4 \\
& PRIMA & 45/2.14/3 & 21/2.10/5 & 0.38 & 54 & 4 & 3 \\
& OIE materials & 1173/4.64/10 & 253/3.51/70 & 0.09 & 119 & 7 & 10 \\
\hline
\end{tabularx}
\end{adjustbox}
\end{table}

\begin{table}[t]
\centering
\caption{Topology metrics of MSE ontologies (Part 2). The columns include Domain, Ontology Name, Absolute/Average/Maximum Depth (Abs./Avg./Max. Depth), Absolute/Average/Maximum Breadth (Abs./Avg./Max. Breadth), Tangledness (Tngld.), Number of Root Classes (NoR), Number of External Classes (NoC), and Number of Leaf Classes (NoL).}
\label{tab:topology-metrics-2}
\begin{adjustbox}{width=\textwidth}
\begin{tabularx}{\textwidth}{m{2cm}|m{3cm}>{\centering\arraybackslash}m{2.5cm}>{\centering\arraybackslash}m{2.5cm}>{\centering\arraybackslash}m{.7cm}*{3}{>{\centering\arraybackslash}m{0.56cm}}}
\hline
\textbf{Domain} & \textbf{Ontology Name} & \textbf{Abs./Avg./Max. Depth} & \textbf{Abs./Avg./Max. Breadth} & \textbf{Tngld.} & \textbf{NoR} & \textbf{NoC} & \textbf{NoL} \\
\hline
\multirow{4}{2cm}{Nano\-materials}
& NPO & 11326/6.45/14 & 1756/3.92/284 & 0.31 & 65 & 284 & 5 \\
& eNanoMapper & 2014/2.60/7 & 775/6.80/99 & 0.00 & 2 & 59 & 5 \\
& NanoMine & 2086/1.19/5 & 1748/46.00/1593 & 0.03 & 1 & 18 & 5 \\
& MaterialsMine & 41/1.58/3 & 26/3.71/13 & 0.00 & 36 & 18 & 5 \\
\hline
\multirow{3}{2cm}{Mechanical Testing}
& EMMO Mechanical Testing & 3403/2.62/7 & 1301/5.56/496 & 0.21 & 13 & 47 & 7 \\
& MOL Brinell & 53/1.43/3 & 37/3.70/23 & 0.00 & 17 & 23 & 3 \\
& MOL Tensile & 415/1.07/3 & 387/32.25/363 & 0.00 & 61 & 11 & 3 \\
\hline
\multirow{3}{2cm}{Additive Manufacturing}
& AMONTOLOGY & 322/1.12/3 & 288/16.94/256 & 0.05 & 5 & 26 & 3 \\
& LPBFO & 695/1.37/4 & 509/11.84/363 & 0.00 & 2 & 33 & 4 \\
& OIE manufacturing & 1304/2.94/8 & 443/6.71/215 & 0.01 & 5 & 9 & 8 \\
\hline
\multirow{2}{2cm}{Batteries}
& BattINFO & 2115/1.13/5 & 1902/35.89/1781 & 0.01 & 10 & 43 & 3 \\
& EMMO BVC & 1757/1.29/ & 1359/18.12/1203 & 0.05 & 11 & 43 & 4 \\
\hline
\multirow{1}{2cm}{Biomaterials}
& DEB & 2135/2.63/5 & 608/5.58/56 & 0.13 & 12 & 601 & 5 \\
\hline
\multirow{1}{2cm}{Sensor}
& SSN & 16/1.00/1 & 16/16.00/16 & 0.00 & 21 & 16 & 3 \\
\hline
\multirow{1}{2cm}{Energy}
& SAREF & 51/2.04/3 & 25/3.57/6 & 0.00 & 35 & 5 & 3 \\
\hline
\end{tabularx}
\end{adjustbox}
\end{table}

\end{document}